\documentclass[onecollarge]{svjour2}       
\smartqed  
\usepackage{graphicx}
\usepackage{color}
\usepackage{csquotes}
\usepackage{enumitem}
\usepackage{url}

\usepackage{mathmacros}
\newcommand{\citep}{\cite}

\newcommand{\vort}{\ensuremath{\zeta}}
\newcommand{\domain}{\ensuremath{\mathcal{D}}}
\newcommand{\sphere}{\ensuremath{\mathbb{S}^2}}
\newcommand{\torus}{\ensuremath{\mathbb{T}^2}}

\newcommand{\macro}[1]{\ensuremath{\mathcal{#1}}} 
\newcommand{\meanfield}[1]{\ensuremath{\mathscr{#1}}}  
\newcommand{\thermo}[1]{\ensuremath{#1}}		       
\newcommand{\partfun}{\ensuremath{\mathcal{Z}}}
\newcommand{\proj}[1]{\mathcal{P}_{#1}}
\DeclareMathOperator{\prob}{Prob}

\renewcommand{\textcolor}[2]{{#2}}

\begin{document}

\title{Additional invariants and statistical equilibria for the 2D Euler equations on a spherical domain
}

\titlerunning{Invariants and equilibria of the 2D Euler equation}        

\author{Corentin Herbert}


\institute{Corentin Herbert \at
              National Center for Atmospheric Research, P.O. Box 3000, Boulder, CO, 80307, USA \\
              Tel.: (303) 497-1636\\
              \email{cherbert@ucar.edu}           
}

\date{Received: date / Accepted: date}

\maketitle

\begin{abstract}
The role of the domain geometry for the statistical mechanics of 2D Euler flows is investigated. It is shown that for a spherical domain, there exists invariant subspaces in phase space which yield additional angular momentum, energy and enstrophy invariants. The microcanonical measure taking into account these invariants is built and a mean-field, Robert-Sommeria-Miller theory is developed in the simple case of the energy-enstrophy measure. The variational problem is solved analytically and a partial energy condensation is obtained. The thermodynamic properties of the system are also discussed.
\end{abstract}


\section{Introduction}\label{introsection}

The tools of equilibrium statistical mechanics have been applied to the study of turbulent flows in a variety of ways. From the pioneering theory of point vortices initiated by Onsager (\cite{Onsager1949}, see also~\cite{Eyink2006}) and developed by many others~\citep{Montgomery1974,Pointin1976,Lundgren1977b,Frohlich1982,Benfatto1987,Campbell1991,Eyink1993,ChorinBook,Kiessling1997,Kiessling2012} to the mean-field theory of Robert, Sommeria and Miller (RSM)~\citep{Robert1991a,Robert1991b,Miller1990} (see also~\cite{MillerThesis,Sommeria1991a,Chavanis1998c,Chavanis2002LNP,MajdaWangBook,Bouchet2012}) through the spectral approach by Kraichnan~\citep{Kraichnan1967,Kraichnan1980}, several theories are available, with their strengths and weaknesses. The models for turbulent flows investigated range from the Euler equations to quasi-geostrophic~\citep{Salmon1976,Frederiksen1980,Merryfield1996,DiBattista1998,Bouchet2002,Venaille2011a,Herbert2012b} or shallow-water equations~\citep{Chavanis2002a,Chavanis2006b}. 
In all these cases, the major feature that statistical mechanics enabled us to better understand is the large-scale organization of the flow. While Onsager's theory gave birth to the early prediction of the existence of negative temperature, and Kraichnan's work yielded the notion of \emph{energy condensation} for fluid flows, it is arguably the RSM theory which provides the most convenient framework to effectively compute the statistical equilibria of the system.
Very often, several statistical equilibria can coexist for a given value of the external parameters, which leads to interesting phase transitions. The long-range nature of the interactions at work in turbulent flows is responsible for new types of phase transitions~\citep{Bouchet2005}. Like with many other systems with long-range interactions~\citep{DauxoisLRIBook,Campa2009}, the energy is not additive in turbulent flows, which has the crucial consequence that the different statistical ensembles may not give equivalent results, even in the thermodynamic limit~\citep{Ellis2000}. Ensemble inequivalence has many manifestations at the thermodynamic level, like the existence of negative specific heats, non concave entropies,...~\citep{Venaille2011a}

For all the above mentioned properties, the domain geometry plays an important role. In the first place, it bears connections with the dynamical invariants of the system, which are the cornerstones of any statistical mechanical approach: for 2D domains, there is an infinity of conserved quantities, the Casimir invariants, because the topology of the domain does not allow for vorticity stretching. These invariants break down for 3D flows, but in some situations, like axisymmetric flows~\citep{Leprovost2006,Naso2010b,Naso2010c}, statistical mechanics methods still provide non-trivial results due to the specific symmetry of the domain. Even for 2D flows, the additional invariants brought about by symmetries of the domain play a crucial role, both at the level of the equilibrium states of the system~\citep{Chavanis1996a,Chen1997,Chavanis1998b,Herbert2012b} or at the level of the thermodynamic properties, for instance through the degeneracies of the eigenvalues of the Laplacian, on the sphere \sphere~\citep{Herbert2012a}, or on the flat torus \torus~\citep{Bouchet2009}. \textcolor{red}{At a more technical level, the geometry of the domain plays a role when solving the mean-field equation which characterizes the statistical equilibria in the Robert-Sommeria-Miller theory. Indeed, in the limit of a linear vorticity--stream function relation, the usual method to solve this equation is to decompose the fields on a basis of Laplacian eigenfunctions. For a rectangular domain, these eigenfunctions are the traditional Fourier modes, while on the sphere~\sphere, for example, they are spherical harmonics, with different properties. For instance the sum of two Laplacian eigenfunctions is still an eigenfunction on a rectangular domain, but it is not so on the sphere~\cite{Bailey1933,Cao1999}. This leads to important differences in the permitted quadratic interactions between different modes~\cite{Tang1978}.}

In this paper, we investigate further the role of the geometry for the RSM theory by focusing on the special case of the sphere~\sphere, due equally to its potential relevance for atmospheric sciences and to its geometric specificities. \textcolor{red}{Numerical simulations of flows on a sphere have been performed in the past~\cite{Tang1978,Williams1978,Basdevant1981,Yoden1993,Dritschel1993,Cho1996}, although there are much less available than for rectangular domains. These simulations show a tendency of the flow to organize spontaneously at the large scales, and some report that the late stage of the flow reduces to simple structures like a solid-body rotation or a quadrupole~\citep{Cho1996,Marston2011}.} Previous computations have shown that the equilibrium states, in an energy-enstrophy theory, are solid-body rotations or dipole flows~\citep{Herbert2012a}. The quadrupole was obtained as a saddle-point of the entropy functional, but it was found to be unstable. Here, we show that it is in fact a statistical equilibrium, by considering the role of new invariants.
We first show that the phase space of Euler flows on a rotating sphere can be decomposed into two invariant subspaces: the evolution of the fundamental (large-scales) modes is independent of the higher-order (smaller-scales) modes, and these modes have an integrable dynamics. We relate this property to the existence of additional angular momentum, energy and enstrophy invariants, and we build a microcanonical measure taking into account these additional invariants. We show that when the microcanonical measure can be approximated by an energy-enstrophy measure, a simple mean-field theory can be formulated, and we solve the corresponding variational problem.

The paper is organized as follows: in section~\ref{eulerinvsection}, we introduce the Euler equations and their standard invariants, and summarize the crucial role played by the invariants (section~\ref{invsection}). The rest of the section is devoted to introducing a decomposition of phase space (section~\ref{phasespacesection}) specific to the spherical geometry, establishing some properties of the interaction tensor (section~\ref{interacttenssection}), related to angular momentum precession (section~\ref{angmomsection}), which result in the appearance of an additional energy invariant (section~\ref{energyinvsection}). We then investigate the consequences of this additional energy invariant for the statistical mechanics of the 2D Euler equations on a spherical domain. We start by showing how the microcanonical measure is built (section~\ref{microcanomeassection}), and solve analytically the variational problem obtained as a limiting case of the general mean-field variational problem associated to the microcanonical measure (section~\ref{linearlimitsection}).

\section{The Invariants of the Euler equations on a 2D rotating sphere}\label{eulerinvsection}

\subsection{The Euler equations and their standard invariants}\label{invsection}

The motion of an inviscid, incompressible fluid over a 2D domain $\domain$ rotating around axis $\vec{\Omega}$ is described by the Euler equations:
\begin{align}
\partial_t \vec{u} + \vec{u} \cdot \nabla \vec{u}&= - \frac 1 {\rho} \nabla P - \proj{\domain} (2 \vec{\Omega} \wedge \vec{u}), \\
\nabla \cdot \vec{u}&=0,
\end{align}
where $\vec{u}$ is the 2D velocity field, $P$ the pressure, $\rho$ the density, and $\proj{\domain} (\vec{x})=\vec{x}-(\vec{x} \cdot \vec{n}) \vec{n}$ is the projection on the tangent plane ($\vec{n}$ being the local normal vector to the domain). Here, we will focus on the case where the domain is a two-dimensional sphere $\domain=\sphere$.
To eliminate the pressure term, it is customary to introduce the (relative) vorticity $\omega = \nabla \wedge \vec{u}$ and the stream function $\omega=-\Delta \psi$. The total vorticity is given by $\zeta=\omega+f$, where the Coriolis parameter $f$ (also called \emph{planetary vorticity}) is given for a 2D sphere by $f=2\Omega \cos \theta$. The Euler equations, expressed in terms of the vorticity, read
\begin{equation}
\partial_t \vort + \vec{u}\cdot \nabla \vort=0.
\end{equation}
They state that the (total) vorticity is conserved along the trajectory of a parcel of fluid. Thus, the vorticity is advected by the flow in a similar manner as a passive scalar. A consequence of this particular form is the existence of the following conserved quantities~\citep{Salmon1988}:
\begin{align}
\macro{E}[\vort]&=\frac 1 2 \int_{\domain} \vec{u}^2 d\vec{r}= \frac 1 2 \int_{\domain} (\zeta-f) \psi d\vec{r},\\
\macro{G}_n[\vort]&= \int_{\domain} \vort^n d\vec{r},
\intertext{where $E$ is the (kinetic) energy of the flow and the moments of the vorticity field $\Gamma_n$ are called \emph{Casimir invariants}. The Casimir invariant corresponding to $n=1$ and $n=2$ are respectively called the \emph{circulation} and the \emph{enstrophy}. Note that in the case of a spherical geometry, the circulation vanishes as a consequence of the Stokes theorem. Hence it will not play any role in the following discussions.
More generally, for any function $g$, the quantity $\int_{\domain} g(\vort)d\vec{r}$ is conserved.
In addition to these invariants, there may be additional conservation laws due to symmetries of the domain. For a 2D rotating sphere $\domain=\sphere$, the \textcolor{red}{vertical component of the} angular momentum,}
\macro{L}_z[\vort]&=\int_{\domain} (\zeta-f) \cos \theta d\vec{r},
\end{align}
is also conserved~\citep{Herbert2012b}.

The consequences of these conservation laws on the physical properties of the flow are numerous and of fundamental importance~\citep{Bouchet2012}: 
\begin{enumerate}[label=(\roman*)]
\item First of all, they are linked (through the degeneracy of the Poisson brackets, see next section) to the existence of an infinity of steady-states for the Euler equations: every flow satisfying $\omega=F(\psi)$ with $F$ an arbitrary function is a stationary solution of the Euler equations. For a spherical domain $\domain=\sphere$, this relation has to be modified to take into account the additional symmetry~\citep{Herbert2012b}: steady-states of the Euler equations are characterized by the relation $\zeta=F(\psi+\Omega_L \cos \theta)$ where $F$ is an arbitrary function and $\Omega_L$ and arbitrary coefficient expressing the fact that there is no preferential reference frame.
\item They can be used to prove the stability of certain classes of such stationary states~\citep{Arnold1966b,Holm1985,Wolansky1998,Ellis2002,Caglioti2007}.
\item They allow for the existence of an \emph{inverse energy cascade}~\citep{FrischBook,Sommeria2001b}. Contrary to 3D turbulence, for which the energy injected at large scales cascades down to the small scales where it is dissipated by viscosity, the reverse occurs for 2D turbulent flows: energy cascades towards the large-scales, while it is enstrophy which cascades towards the small-scales~\citep{Fjortoft1953,Kraichnan1967,Leith1968,Batchelor1969}.
\item As suggested by the existence of an inverse cascade, without a mechanism of dissipation for the energy at large scales, the energy piles up at the largest scales. Spontaneous organization of 2D flows in large-scale coherent structures has indeed been observed~\citep{Tabeling2002}, both numerically~\citep{McWilliams1984,Santangelo1989} and experimentally~\citep{Couder1986,NguyenDuc1988,vanHeijst1989a,vanHeijst1989b}.
\item They are responsible for the relative success of equilibrium statistical mechanics in predicting the emergence of the large-scale coherent structures~\citep{ChorinBook,Chavanis1996phd,Sommeria2001b,BouchetThesis,MajdaWangBook,VenailleThesis,CorvellecThesis,HerbertThesis,Bouchet2012}. By contrast, equilibrium statistical mechanics yields trivial results in the case of 3D turbulent flows~\citep{Lee1952,Pomeau1995}.
\item They play an important role in the proof of the validity of a mean-field statistical theory~\citep{Michel1994b,Boucher1999,Robert2000,Bouchet2010,Bouchet2012}, like the Robert-Sommeria-Miller theory~\citep{Miller1990,Robert1991a,Robert1991b}.
\end{enumerate}

\subsection{Notations and Phase-space decomposition}\label{phasespacesection}

In the vorticity form, the 2D Euler equations can be recast as
\begin{equation}
\partial_t \vort + \{ \psi, \vort\}=0, \quad \vort=\omega+f, \omega=-\Delta\psi, \quad \text{with } \vort,\psi \in L^2(\domain),
\end{equation}
where the Poisson bracket $\{A,B\}=\partial_x A \partial_y B -\partial_x B \partial_y A$ endows $L^2(\domain)$ with a Lie algebra structure: it is bilinear: $\{\alpha A + \beta B, C\}= \alpha \{A,C\}+\beta \{ B, C\}$, antisymmetric: $\{A,B \}=-\{B, A\}$, and satisfies the Jacobi identity
\begin{equation}
\{ \{ A, B\}, C \} + \{ \{ B, C\}, A \} + \{ \{ C, A\}, B \} = 0.
\end{equation}
Besides, if $F$ is any differentiable function, $\{A,F(A)\}=0$. Note that the degeneracy of the Poisson bracket has very important consequences: it is responsible for the existence of an infinite number of conserved quantities, the Casimir invariants, introduced in section~\ref{invsection}, and for the existence of an infinite number of steady-states for the Euler equations~\citep{Salmon1988,Morrison1998}.

Given the relation between vorticity and stream function, it is natural to introduce the eigenfunctions of the Laplacian. On the sphere $\domain=\sphere$, the eigenfunctions of the Laplacian are the spherical harmonics $Y_{nm}$, with $n \in \N, -n \leq m \leq n$, and the corresponding eigenvalues are given by $-\lambda_n=-n(n+1)$, so that $\Delta Y_{nm}=-\lambda_n Y_{nm}$. Note that the eigenvalues are degenerate: for $n \in \N$, there are $2n+1$ eigenfunctions $Y_{nm}$ with eigenvalue $-\lambda_n$, for $-n \leq m \leq n$. Let us note $V_n$ the corresponding eigenspace: $\displaystyle V_n=\Ker(\Delta+\lambda_n \Id)=\Vect_{-n\leq m \leq n} Y_{nm}$. Hence, we have $\dim V_n=2n+1$. We can decompose the phase-space in the following manner:
\begin{align}
L^2(\sphere)= \bigoplus_{n=0}^{+\infty} V_n&=V_1 \oplus V_{\infty}, \quad \text{ with } V_{\infty}= \bigoplus_{n \neq 1} V_n,\label{phasespdeceq}
\intertext{so that the vorticity and stream function fields can be decomposed as}
\vort&=\vort_1+\vort_{\infty},\\
\psi&=\psi_1+\psi_{\infty}, \quad\text{with } \vort_1=-\Delta \psi_1+f, \vort_{\infty}=-\Delta \psi_{\infty}.
\end{align}
In fact, since the spherical harmonics form an orthonormal basis of the Hilbert space $L^2(\sphere)$, the direct sums in Eq. \ref{phasespdeceq} are orthogonal sums for the standard $L^2$ scalar product:
\begin{align}
\langle f | g \rangle&=\int_0^{2\pi} d\phi \int_0^{\pi} \sin \theta d\theta f(\theta,\phi) g(\theta,\phi)^*,\\
\langle Y_{nm} | Y_{pq} \rangle &= \delta_{np} \delta_{mq},\\
\forall n\neq n' \in \N, V_{n'} &\subset V_n^{\perp}.
\end{align}
In particular, $\langle \vort_1 | \vort_{\infty} \rangle = \langle \psi_1 | \psi_\infty\rangle = \langle \psi_1 | \vort_\infty\rangle = \langle \vort_1 | \psi_\infty\rangle = 0$.
It is useful to introduce the orthogonal projections on each eigenspace $V_n$:
\begin{equation}
\proj{n} f = \sum_{m=-n}^n \langle f | Y_{nm} \rangle Y_{nm}, \quad \proj{\infty} = \bigoplus_{n=2}^{+\infty} \proj{n}.
\end{equation}
Hence, $\psi_1 = \proj{1} \psi, \vort_1 = \proj{1} \vort$, $\psi_\infty = \proj{\infty} \psi$ and $\vort_\infty = \proj{\infty} \vort$. Clearly, $\proj{n}^2=\proj{n}, \proj{\infty}^2=\proj{\infty}$. 

Projecting the 2D Euler equations gives
\begin{align}
\partial_t \vort_1 + \proj{1} \{\psi,\omega\}+\proj{1}\{\psi,f\}&= 0,\\
\partial_t \vort_{\infty} + \proj{\infty} \{\psi,\vort\}&= 0.
\end{align}
Note that the Coriolis parameter $f=2\Omega\cos \theta$ is proportional to $Y_{10}$: $f=f_{10}Y_{10}$. In particular, $f \in V_1$ and we will show that $\proj{1}\{\psi,f\}=\{\psi_1,f\}$ (see section~\ref{energyinvsection}).
We will also show that the nonlinear interaction term $\{\psi,\omega\}$ has a vanishing projection on the fundamental modes: $\proj{1} \{\psi,\omega\}=0$ (section~\ref{energyinvsection}).

To understand the interactions between various $V_n$ \enquote{shells}, it is useful to write the time evolution of each mode. Decomposing the vorticity $\vort$ in the basis of Laplacian eigenfunctions $Y_{nm}$, we have
\begin{align}
&\dot{\vort}_{nm} + \sum_{n_1,m_1} \sum_{n_2,m_2} \psi_{n_1 m_1} A^{nm}_{n_1m_1n_2m_2} \vort_{n_2 m_2}=0,\\
\intertext{where}
&\vort(\theta,\phi)=\sum_{n,m} \vort_{nm} Y_{nm}(\theta,\phi),\quad \psi(\theta,\phi)=\sum_{n,m} \psi_{nm} Y_{nm}(\theta,\phi),\\
\intertext{and}
&A^{nm}_{n_1m_1n_2m_2}=\langle \{ Y_{n_1m_1}, Y_{n_2m_2} \} | Y_{nm}\rangle,
\end{align}
is the interaction tensor. 
The condition for the vorticity and stream function to take real values reads $\vort_{n,-m}=(-1)^m\vort_{nm}^*, \psi_{n,-m}=(-1)^m\psi_{nm}^*$.
The antisymmetric character of the Poisson brackets carries over to the interaction tensor:
\begin{equation}
A^{nm}_{n_1m_1n_2m_2}=-A^{nm}_{n_2m_2n_1m_1}.
\end{equation}
Note that we have $\vort_{nm}=\lambda_n \psi_{nm}+\delta_{n1}\delta_{m0}f_{10}$. Hence the evolution of a given $(n,m)$ mode is determined by two contributions: a quadratic term $\mathcal{Q}_{n,m}[\vort]$ corresponding to self-interaction of the vorticity field, and a linear term $\mathcal{L}_{n,m}[\vort]$ corresponding to interactions between relative vorticity and planetary vorticity (due to rotation):
\begin{align}
\dot{\vort}_{nm}&=-\mathcal{Q}_{n,m}[\vort]+\mathcal{L}_{n,m}[\vort],\\
\mathcal{Q}_{n,m}[\vort]&=\sum_{n_1,m_1} \sum_{n_2,m_2} \vort_{n_1 m_1} \frac{ A^{nm}_{n_1m_1n_2m_2}}{\lambda_{n_1}} \vort_{n_2 m_2},\\
\mathcal{L}_{n,m}[\vort]&=\sum_{n_2,m_2} \vort_{n_2 m_2} \frac{A^{nm}_{10n_2m_2}}{\lambda_{1}} f_{10}.
\end{align}
The quadratic term involves the contraction of an symmetric tensor $\vort_{n_1 m_1}\vort_{n_2 m_2}$ with an arbitrary tensor ${ A^{nm}_{n_1m_1n_2m_2}}/{\lambda_{n_1}}$; only the symmetric part of the latter subsists. Let us denote it $B^{nm}_{n_1m_1n_2m_2}$. As $A^{nm}_{n_1m_1n_2m_2}$ is antisymmetric, it is simply given by
\begin{equation}
B^{nm}_{n_1m_1n_2m_2} = \frac 1 2 \left(\frac{ A^{nm}_{n_1m_1n_2m_2}}{\lambda_{n_1}}+\frac{ A^{nm}_{n_2m_2n_1m_1}}{\lambda_{n_2}}\right)= \frac{A^{nm}_{n_1m_1n_2m_2}}{2} \left(\frac {1}{\lambda_{n_1}}-\frac{1}{\lambda_{n_2}}\right),
\end{equation}
and we have
\begin{equation}
\mathcal{Q}_{n,m}[\vort]=\sum_{n_1,m_1} \sum_{n_2,m_2} \vort_{n_1 m_1}  B^{nm}_{n_1m_1n_2m_2} \vort_{n_2 m_2}.
\end{equation}
The interaction tensor $B^{nm}_{n_1m_1n_2m_2}$ describes the interactions of a \emph{triad} of \emph{wave vectors} $(n,m),(n_1,m_1),(n_2,m_2)$. The analogous quantity in the case of a planar geometry can easily be computed. In particular, the standard \emph{selection rule} states that a given triad can interact only if their wave vectors $\vec{p},\vec{q},\vec{k}$ form a triangle: $\vec{p}+\vec{q}=\vec{k}$. In the case of a spherical domain, the selection rules are not as simple.
 Nevertheless, some properties of the interaction tensor can be analytically derived in an elementary way. We present some of these properties in the following section.

\subsection{Properties of the interaction tensor}\label{interacttenssection}
\label{interactgensection}
As usual in two-dimensional turbulence~\citep{Kraichnan1980}, energy and enstrophy conservation can be expressed in terms of the symmetric interaction tensor:
\begin{align}
 \frac{B^{nm}_{n_1m_1n_2m_2}}{\lambda_n} +  \frac{B^{n_1m_1}_{n_2m_2nm}}{\lambda_{n_1}} + \frac{B^{n_2m_2}_{nmn_1m_1}}{\lambda_{n_2}} &=0,\\
 B^{nm}_{n_1m_1n_2m_2} + B^{n_1m_1}_{n_2m_2nm} + B^{n_2m_2}_{nmn_1m_1} &=0.
\end{align}
Recalling the expression for the spherical harmonics~\citep{GradshteynRyzhik} (see also Appendix \ref{sphericalharmsection} for a summary of the properties of the spherical harmonics and associated Legendre polynomials used here),
\begin{equation}
Y_{nm}(\theta,\phi)= c_{nm} P_n^m (\cos \theta) e^{im\phi},\quad c_{nm} = \sqrt{\frac {2n+1} {4\pi} \frac{(n-m)!}{(n+m)!}},
\end{equation}
the Poisson brackets $\{ Y_{n_1m_1},Y_{n_2m_2}\}$ reads
\begin{align}
\{ Y_{n_1m_1},Y_{n_2m_2}\}&= \frac {\partial Y_{n_1m_1}}{\partial \phi} \frac {\partial Y_{n_2m_2}}{\partial x} - \frac {\partial Y_{n_1m_1}}{\partial x}\frac {\partial Y_{n_2m_2}}{\partial \phi},\\
&=i c_{n_1m_1} c_{n_2m_2} \lbrack m_1 P_{n_1}^{m_1}(x){P_{n_2}^{m_2}}'(x)-m_2P_{n_2}^{m_2}{P_{n_1}^{m_1}}'(x)\rbrack e^{i(m_1+m_2)\phi},
\end{align}
where $x=\cos \theta$. Multiplying by $Y_{nm}^*$ and integrating on the sphere, we obtain
\begin{equation}\label{Atensoreq}
A^{nm}_{n_1m_1n_2m_2}= 2i\pi c_{nm}c_{n_1m_1}c_{n_2m_2} \delta_{m_1+m_2,m} \lbrack m_1 I^{n_2m_2}_{nmn_1m_1}-m_2 I^{n_1m_1}_{nmn_2m_2}\rbrack,
\end{equation}
where we have defined the \emph{structure integrals}
\begin{equation}
I^{n_2m_2}_{nmn_1m_1} = \int_{-1}^1 P_n^m(x) P_{n_1}^{m_1}(x){P_{n_2}^{m_2}}'(x) dx,
\end{equation}
and used the property
\begin{equation}
\int_0^{2\pi} e^{it\phi}d\phi=2\pi \delta_{t0}.
\end{equation}
Note that in~\eqref{Atensoreq}, it is explicit that the tensor $A^{nm}_{n_1m_1n_2m_2}$ is antisymmetric, as the quantity in the brackets is and the prefactors are symmetric. Moreover, the symmetry property $I^{n_2m_2}_{nmn_1m_1}=I^{n_2m_2}_{n_1m_1nm}$ holds, and it is also clear that $A^{nm}_{n_10n_20}=0$, which means that two purely zonal modes do not interact.

In full generality, the individual interaction between two modes $(n_1,m_1),(n_2,m_2)$ belonging to the same \emph{shell} --- $n_1=n_2$ --- is nonzero. Nevertheless, the sum of all the contributions from the same shell vanish:
\begin{equation}
\sum_{m_1,m_2}  \vort_{n_1m_1} \frac{A^{nm}_{n_1m_1n_2m_2}}{\lambda_{n_2}} \vort_{n_2m_2}=0,
\end{equation}
because we are contracting a symmetric tensor with an antisymmetric one. Even more evidently, $B^{nm}_{n_1m_1n_2m_2} =0$ when $n_1=n_2$.

As a consequence, the summations in $\mathcal{Q}_{n,m}[\vort]$ and $\{\psi,\omega\}$ have no $n_1=n_2$ contribution and no $m_1=m_2=0$ contribution:
\begin{align}
\mathcal{Q}_{n,m}[\vort]&=\sum_{n_1 \neq n_2} \sum_{(m_1,m_2) \neq (0,0)} \vort_{n_1 m_1}  B^{nm}_{n_1m_1n_2m_2} \vort_{n_2 m_2},\\
\{ \psi, \omega \} &=\sum_{n,m} \mathcal{Q}_{n,m}[\vort] Y_{nm},\\
 &= \sum_{n,m} \left( \sum_{n_1 \neq n_2} \sum_{(m_1,m_2) \neq (0,0)}  \vort_{n_1m_1} B^{nm}_{n_1m_1n_2m_2} \vort_{n_2m_2}\right) Y_{nm}.
\end{align}

\textcolor{red}{Direct computations (shown in Appendix \ref{directcompusection}) allow to show that for any $n_1,n_2 \in \N^*$, and $-n_1 \leq m_1 \leq n_1$, $- n_2 \leq m_2 \leq n_2$, we have:
\begin{enumerate}[label=(\roman*)]
\item For $-1 \leq m \leq 1$, $B_{n_1m_1n_2m_2}^{1m}=0$ --- the fundamental modes cannot be excited by any quadratic interaction: $\mathcal{Q}_{1,m}[\vort]=0$.
\item If $n_1 > 1$, $A_{n_1m_110}^{1m}=0$ --- the fundamental modes cannot be excited by rotation-vorticity interactions as soon as the vorticity modes are not fundamental themselves (this is a direct consequence of the previous point).
\item For $n \in \N^*, n_1=1, n_2=1, -n \leq m \leq n$, $B_{1m_11m_2}^{nm}=0$ --- quadratic interactions of fundamental modes vanish.
\item If in addition $n>1$, $A_{1m_11m_2}^{nm}=0$ --- rotation-vorticity interactions with a fundamental vorticity mode vanish.
\end{enumerate}
The first two points can be obtained without direct computations, as they are related to angular momentum conservation, as we show in the next section.}

\textcolor{red}{\subsection{The angular momentum invariants}\label{angmomsection}}

\textcolor{red}{\subsubsection{Angular momentum precession}}

\textcolor{red}{Let us consider the global angular momentum of the flow, in the non-rotating reference frame $\mathcal{R}_0$:
\begin{align} 
\vec{\macro{L}}_0[\vort]&=\int_\domain \vec{e_r} \wedge \vec{u} d\vec{r},\\
\vec{\macro{L}}_0[\vort]&= \int_\domain \vort \vec{e_r}d\vec{r}.
\intertext{Note that in a reference frame $\mathcal{R}'$ rotating with angular velocity $\Omega' \vec{e_z}$, the angular momentum would be $\vec{\macro{L}}'=\vec{\macro{L}}_0-\frac {8\pi}{3}\Omega' \vec{e_z}$. In particular, in the reference frame $\mathcal{R}$ rotating with the sphere ($\Omega'=\Omega$),}
\vec{\macro{L}}[\vort]&=\int_\domain (\vort-f) \vec{e_r}d\vec{r}.
\end{align}
A simple calculation shows that in an arbitrary rotating reference frame $\mathcal{R}'$, we have
\begin{align}
\frac{d \vec{\macro{L}}'}{dt} \Big|_{\mathcal{R}'}&=\frac{d \vec{\macro{L}}_0}{dt} \Big|_{\mathcal{R}'}=2\Omega' \int_\domain \cos \theta \vec{u} d\vec{r},
\intertext{so that in particular, the angular momentum is conserved in the non-rotating reference frame $\mathcal{R}_0$. As a consequence, the motion of the angular momentum in a rotating reference frame, and in particular in the reference frame $\mathcal{R}$ rotating with the sphere, is a precession motion:}
\frac{d \vec{\macro{L}}}{dt} \Big|_{\mathcal{R}}&=-\Omega \vec{e_z} \wedge \vec{\macro{L}}.
\end{align}
It follows from this that both the vertical component of the angular momentum $\macro{L}_z[\vort]$ and the norm $\macro{L}^2[\vort]=\| \vec{\macro{L}}[\vort] \|_2^2=\macro{L}_x[\vort]^2+\macro{L}_y[\vort]^2+\macro{L}_z[\vort]^2$ are conserved:
\begin{align}
\frac{d}{dt} \macro{L}_z[\vort] = \frac{d}{dt} \macro{L}^2[\vort] = 0.
\end{align}}

\textcolor{red}{\subsubsection{Integrability of the dynamics in $V_1$}\label{integrabilitysection}}

\textcolor{red}{The existence of the additional invariant $\macro{L}^2[\vort]$ has important consequences on the dynamics. The three components of the angular momentum in the rotating reference frame $\mathcal{R}$ can be expressed as scalar products of the vorticity field with fixed vectors in $L^2(\domain)$:
\begin{align}
\macro{L}_x[\vort]&=\int_\domain (\vort-f) \sin \theta \cos \phi d\vec{r}= \sqrt{\frac{2\pi}{3}} \langle \vort-f | Y_{1,-1}-Y_{11} \rangle,\\
\macro{L}_y[\vort]&=\int_\domain (\vort-f) \sin \theta \sin \phi d\vec{r}= -i \sqrt{\frac{2\pi}{3}} \langle \vort-f | Y_{1,-1}+Y_{11} \rangle,\\
\macro{L}_z[\vort]&=\int_\domain (\vort-f) \cos \theta d\vec{r}= \sqrt{\frac{4\pi}{3}} \langle \vort-f | Y_{10} \rangle.
\intertext{Decomposing the vorticity field on the orthonormal basis made by the spherical harmonics, $\omega(\theta,\phi)=\sum_{n,m}\omega_{nm}Y_{nm}(\theta,\phi)$, we have}
\macro{L}_x[\vort]&=\sqrt{\frac{2\pi}{3}}(\omega_{1,-1}-\omega_{11})=\sqrt{\frac{8\pi}{3}} \Re(\omega_{1,-1}),\\
\macro{L}_y[\vort]&=-i\sqrt{\frac{2\pi}{3}}(\omega_{1,-1}+\omega_{11})=\sqrt{\frac{8\pi}{3}} \Im(\omega_{1,-1}),\\
\macro{L}_z[\vort]&=\sqrt{\frac{4\pi}{3}}\omega_{10},
\end{align}
where we have used $\omega_{11}=-\omega_{1,-1}^*$. We see that the three components of angular momentum completely determine the coefficients $\omega_{10}$, $\omega_{11}$ and $\omega_{1,-1}$, or in other words, they completely determine $\omega_1=\proj{1} \omega$. Besides, the dynamics of the angular momentum is known: it is simply a precession motion. Hence, we can easily obtain the evolution equations for these coefficients:
\begin{align}
\begin{cases}
\dot{\omega}_{10}&=0\\
\dot{\omega}_{11}&=i\Omega \omega_{11}\\
\dot{\omega}_{1,-1}&=-i\Omega \omega_{1,-1}
\end{cases}\label{vonedyneq}
\end{align}
This set of equations makes it clear that:
\begin{enumerate}[label=(\roman*)]
\item The evolution of $\omega_1$ does not depend on the other modes.
\item In terms of the interaction tensor, we have for any $m,n_1,m_1,n_2,m_2$, $B_{n_1m_1n_2m_2}^{1m}=0$ and $A_{n_1m_110}^{10}=0$, $A_{n_1m_110}^{11}=i\Omega\delta_{n_11}\delta_{m_11}, A_{n_1m_110}^{1,-1}=-i\Omega\delta_{n_11}\delta_{m_1,-1}$.
\item The dynamics in $V_1$ is integrable. Indeed, \eqref{vonedyneq} is easily solved, and with $\omega_{1,-1}(0)=re^{-i\phi_0}$, we must have $r^2=3(\macro{L}_x[\vort]^2+\macro{L}_y[\vort]^2)/8\pi$, and
\begin{align}\label{fundmodesdyneq}
\omega_1(\theta,\phi,t) &= \frac{3}{4\pi}\Big\lbrack\macro{L}_z[\vort] \cos \theta+ \sqrt{\macro{L}^2[\vort]-\macro{L}_z[\vort]^2}\sin \theta \cos(\Omega t + \phi+\phi_0)\Big\rbrack.
\end{align}
\end{enumerate}
}

\subsection{Projected dynamics and additional energy invariant}\label{energyinvsection}

From the previous discussion, we know that we always have $B^{1m}_{n_1m_1n_2m_2} = 0$ --- in particular, if $n_1 \neq n_2$, $A^{1m}_{n_1m_1n_2m_2} = 0$. As a consequence, for $-1 \leq m \leq 1$, $\mathcal{Q}_{1,m}[\vort]=0$, and $\{ \psi, \omega \}$ has no component in $V_1$: 
\begin{equation}
\displaystyle \proj{1} \{ \psi, \omega \} = \sum_{m=-1}^1 \mathcal{Q}_{1,m}[\vort] Y_{1m}=0.
\end{equation}
As a consequence, the projected dynamics read
\begin{align}
\partial_t \vort_1 +\proj{1}\{\psi,f\}&= 0,\\
\partial_t \vort_{\infty} + \proj{\infty} \{\psi,\vort\}&= 0.
\end{align}
In addition, $\mathcal{L}_{1,m}[\vort]=\mathcal{L}_{1,m}[\vort_1]$, and
\begin{align}
\proj{1}\{\psi,f\}&=\sum_{m=-1}^1 \mathcal{L}_{1,m}[\vort_1] Y_{1m}=\proj{1} \{\psi_1,f\}.
\end{align}
As we also have $A^{nm}_{1m_11m_2} = 0$ if $n>1$ (section \ref{fundmodeintersec}), $\{\psi_1,f\} \in V_1$ and $\proj{1} \{\psi_1,f\}=\{\psi_1,f\}$. Finally, the projected dynamics is
\begin{align}
\partial_t \vort_1 +\{\psi_1,f\}&= 0,\\
\partial_t \vort_{\infty} + \{\psi_1,\omega_{\infty}\}+ \{\psi_{\infty},\omega_1\}+\{\psi_{\infty},\omega_{\infty}\}+\{\psi_{\infty},f\}&= 0.
\end{align}

In this form, it is clear that the evolution of the fundamental modes (i.e. the largest scales for the vorticity, $\vort_1$) does not depend on the other modes (i.e. all the other scales for the vorticity, $\vort_{\infty}$), as already shown by the angular momentum conservation arguments. The converse, however, is not true. Besides, the dynamics of the fundamental modes is integrable: see \eqref{fundmodesdyneq}.

The way the original dynamical system splits up into two parts, one of which evolves independently of the other, has consequences at the level of conserved quantities. Indeed, recall that $\macro{E}[\vort]=-\langle \omega | \psi \rangle / 2=-\langle \omega_1 | \psi_1\rangle/2 -\langle \omega_{\infty} | \psi_{\infty}\rangle /2$ by orthogonality properties, so that
\begin{align}
\macro{E}[\vort]&=\macro{E}[\vort_1]+\macro{E}[\vort_{\infty}].
\intertext{Similarly, $\macro{L}_z[\vort]=\macro{L}_z[\vort_1]$, and $\macro{L}^2[\vort]=\macro{L}^2[\vort_1]$. The enstrophy also splits up into two separate contributions:}
\macro{G}_2[\vort]&=\macro{G}_2[\vort_1]+\macro{G}_2[\vort_{\infty}],\label{enstrophyspliteq}
\end{align}
because $\macro{G}_2[\vort]=\| \vort \|^2$ where $\|\cdot \|$ is the hermitian norm, so that~\eqref{enstrophyspliteq} is just the Pythagorean theorem. In general, the Casimir invariants $\macro{G}_n[\vort]$ with $n \geq 3$ do not split into separate contributions from $\vort_1$ and $\vort_{\infty}$.
It follows from these decompositions that $\macro{E}[\vort_1]$ and $\macro{E}[\vort_{\infty}]$ are conserved independently (i.e. not only the sum is conserved). \textcolor{red}{An explicit proof is given in \ref{energyinvdirproofsection}, but it also follows from the angular momentum conservation properties. In fact, the invariants are not independent: we have $\macro{E}[\vort_1]=\frac{3}{16\pi}\macro{L}^2[\vort]$.}

As explained above, there exist interactions between fundamental modes ($n=1$) and other modes ($n\geq 2$), but they cannot transfer energy into the shell $V_1$, and thus conserve $\macro{E}[\vort_{\infty}]$ as a whole.
For a given triad $(1,m),(n_1,m_1),(n_2,m_2)$, global enstrophy conservation reads $B^{1m}_{n_1m_1n_2m_2} + B^{n_1m_1}_{n_2m_21m} + B^{n_2m_2}_{1mn_1m_1} =0$, but as $B^{1m}_{n_1m_1n_2m_2} = 0$, it remains that $B^{n_1m_1}_{n_2m_21m} + B^{n_2m_2}_{1mn_1m_1}=0$, which means that energy can only be exchanged between shells $V_{n_1}$ and $V_{n_2}$ in such a way that $\macro{E}[\vort_{n_1}]+\macro{E}[\vort_{n_2}]$ is constant (and in particular, so is $\macro{E}[\vort_{\infty}]$).

\section{Statistical mechanics with an additional energy invariant}

\subsection{The microcanonical measure}\label{microcanomeassection}

\subsubsection{The Liouville theorem}\label{liouvillesection}

The cornerstone of equilibrium statistical mechanics is the Liouville theorem, which states that the volume in phase space is conserved by the dynamics, or in other words, that the Lebesgue measure is invariant under the flow of the dynamical system. For canonical Hamiltonian flows, this is a trivial property. When the phase space does not have such a canonical structure, one should verify the Liouville theorem by hand. Here, we have,
\begin{align}
\frac{\partial \dot{q}_{nm}}{\partial q_{n'm'}}&=-\frac{\partial \mathcal{Q}_{nm}[\vort]}{\partial q_{n'm'}}+\frac{\partial \mathcal{L}_{nm}[\vort]}{\partial q_{n'm'}},\\
&=-2 \sum_{n_1,m_1}  B^{nm}_{n'm'n_1m_1}\vort_{n_1m_1} + A_{10n'm'}^{nm} \frac{f_{10}}{\lambda_1}.
\intertext{In particular, when $(n',m')=(n,m)$,}
\frac{\partial \dot{q}_{nm}}{\partial q_{nm}}&=-2 \sum_{n_1,m_1}  B^{nm}_{nmn_1m_1}\vort_{n_1m_1} + A_{10nm}^{nm} \frac{f_{10}}{\lambda_1}.
\intertext{One can prove that $B^{nm}_{nmn_1m_1}=0$, using for instance the enstrophy conservation relation $ B^{nm}_{n_1m_1n_2m_2} + B^{n_1m_1}_{n_2m_2nm} + B^{n_2m_2}_{nmn_1m_1} =0$. Imposing $(n_2,m_2)=(n,m)$ in this relation and using the fact that $B^{n_1m_1}_{n_2m_2nm}$ vanishes because $n_2=n$, we have $B^{nm}_{n_1m_1nm} + B^{nm}_{nmn_1m_1} =0$, and using the symmetry property with respect to the exchange of the bottom couples of indices, we obtain $B^{nm}_{nmn_1m_1}=0$. It remains}
\frac{\partial \dot{q}_{nm}}{\partial q_{nm}}&= A_{10nm}^{nm} \frac{f_{10}}{\lambda_1}.
\end{align}
In general, $A_{10nm}^{nm}$ is not zero, which means that $\frac{\partial \dot{q}_{nm}}{\partial q_{nm}}$ is not zero in the presence of rotation. However, the Liouville theorem is a weaker property: it only states that
\begin{align}
\sum_{n,m }\frac{\partial \dot{q}_{nm}}{\partial q_{nm}}&=0.
\intertext{When each term vanishes individually, we say that the system satisfies a \emph{detailed Liouville theorem}: $\frac{\partial \dot{q}_{nm}}{\partial q_{nm}}=0$. Let us introduce a property which is stronger than the Liouville theorem but weaker than the detailed Liouville theorem: we say that the system satisfies a \emph{semi-detailed Liouville theorem} if for each $n$,}
\sum_{m=-n}^n\frac{\partial \dot{q}_{nm}}{\partial q_{nm}}&=0.
\intertext{In the case of the 2D Euler equations on a rotating sphere, we have}
\sum_{m=-n}^n\frac{\partial \dot{q}_{nm}}{\partial q_{nm}}&= \frac{f_{10}}{\lambda_1}\sum_{m=-n}^n A_{10nm}^{nm},
\end{align}
and since 
\begin{align}
A_{10nm}^{nm}&=-2i\pi c_{nm}^2c_{10}mI_{nmnm}^{10},\\
&=-2i\pi \sqrt{\frac{3}{4\pi}}\frac{2n+1}{4\pi} \times m \frac{(n-m)!}{(n+m)!}\underbrace{\int_{-1}^1 P_n^m(x)P_n^m(x)dx}_{\frac{2(n+m)!}{(2n+1)(n-m)!}},\\
&=-i \sqrt{\frac{3}{4\pi}} m,
\end{align}
and clearly, $\displaystyle \sum_{m=-n}^n m =0$. As a conclusion, the detailed Liouville theorem holds for the 2D Euler equations on a spherical domain, but it breaks in the presence of rotation. However, in that case, a semi-detailed Liouville theorem is still valid. In both cases, the Liouville theorem holds, and it does so for every truncation which is symmetric in the azimuthal number $m$.

\subsubsection{Construction of the microcanonical measure}

The statistical properties of the system (in the limit of large times) are described by an invariant measure. A standard choice of invariant measure in equilibrium statistical mechanics is that of the microcanonical measure. The microcanonical measure is absolutely continuous with respect to Lebesgue, with a density which depends only on the dynamical invariants of the system. It is therefore an invariant measure as a consequence of the Liouville theorem. For a system such as the 2D Euler equations, the difficulties are twofold: first, the phase-space is infinite-dimensional, and second, there is an infinite number of conserved quantities. Following \cite{Bouchet2010}, we can build the microcanonical measure as a limit measure of finite-dimensional approximations.

Let us fix $N,K>0$ and consider the projections $\proj{i,N}=\proj{i}\oplus\cdots\oplus \proj{N}$, and the finite-dimensional finite-moments measure
\begin{align}
\mu_{\thermo{E_1},\thermo{E}_\infty,\thermo{L}_z,\thermo{\Gamma}_1,\cdots,\thermo{\Gamma}_K}^N(d\vort)&= d(\proj{1,N}\vort)\frac{\delta(\macro{E}[\vort_1]-\thermo{E}_1)\delta(\macro{E}[\proj{2,N}\vort]-\thermo{E}_\infty) \delta(\macro{L}_z[\vort_1]-\thermo{L}_z)\prod_{k=1}^K \delta(\macro{G}_k[\proj{1,N}\vort]-\thermo{\Gamma}_k)}{\Omega^N(\thermo{E_1},\thermo{E}_\infty,\thermo{L}_z,\thermo{\Gamma}_1,\cdots,\thermo{\Gamma}_K)} ,\\
&=\prod_{n=1}^N\prod_{m=-n}^nd\vort_{nm}\frac{\delta(\macro{E}[\vort_1]-E_1)\delta(\macro{E}[\proj{2,N}\vort]-E_\infty) \delta(\macro{L}_z[\vort_1]-\thermo{L}_z)\prod_{k=1}^K \delta(\macro{G}_k[\proj{1,N}\vort]-\thermo{\Gamma}_k)}{\Omega^N(\thermo{E_1},\thermo{E}_\infty,\thermo{L}_z,\thermo{\Gamma}_1,\cdots,\thermo{\Gamma}_K)},
\intertext{with}
\macro{E}[\proj{2,N}\vort]&=\sum_{n=2}^N \sum_{m=-n}^n \frac {|\vort_{nm}|^2}{2\lambda_n}, \\
 \macro{G}_k[\proj{1,N}\vort]&=\sum_{n_1=1}^N \sum_{m_1=-n_1}^{n_1}\cdots \sum_{n_k=1}^N \sum_{m_k=-n_k}^{n_k} \vort_{n_1m_1}\cdots \vort_{n_km_k} \int_\domain Y_{n_1m_1}(\theta,\phi)\cdots Y_{n_km_k}(\theta,\phi) \sin \theta d\theta d\phi,
\intertext{and where}
\Omega^N(\thermo{E_1},\thermo{E}_\infty,\thermo{L}_z,\thermo{\Gamma}_1,\cdots,\thermo{\Gamma}_K)&=\int \prod_{n=1}^N\prod_{m=-n}^nd\vort_{nm} \delta(\macro{E}[\vort_1]-\thermo{E}_1)\delta(\macro{E}[\proj{2,N}\vort]-\thermo{E}_\infty) \delta(\macro{L}_z[\vort_1]-\thermo{L}_z)\prod_{k=1}^K \delta(\macro{G}_k[\proj{1,N}\vort]-\thermo{\Gamma}_k)
\end{align}
is the finite-dimensional finite-moments \emph{structure function}~\citep{KhinchinBook}. The microcanonical limit measure is defined as the limit measure
\begin{align}
\mu_{\thermo{E_1},\thermo{E}_\infty,\thermo{L}_z,\thermo{\Gamma}_1,\ldots}&=\lim_{K\to +\infty}\lim_{N\to \infty} \mu_{\thermo{E_1},\thermo{E}_\infty,\thermo{L}_z,\thermo{\Gamma}_1,\ldots,\thermo{\Gamma}_K}^N,
\intertext{which we simply denote as}
\mu_{\thermo{E_1},\thermo{E}_\infty,\thermo{L}_z,\thermo{\Gamma}_1,\ldots}(d\vort)&=d\vort\frac{\delta(\macro{E}[\vort_1]-\thermo{E}_1)\delta(\macro{E}[\vort_{\infty}]-\thermo{E}_\infty) \delta(\macro{L}_z[\vort_1]-\thermo{L}_z)\prod_{k=1}^{+\infty} \delta(\macro{G}_k[\vort]-\thermo{\Gamma}_k)}{\Omega(\thermo{E_1},\thermo{E}_\infty,\thermo{L}_z,\thermo{\Gamma}_1,\cdots)},
\end{align}
where $\Omega(\thermo{E_1},\thermo{E}_\infty,\thermo{L}_z,\thermo{\Gamma}_1,\cdots)$ is the \emph{structure function}. Although the finite-dimensional finite-moments measures $\mu_{\thermo{E_1},\thermo{E}_\infty,\thermo{L}_z,\thermo{\Gamma}_1,\ldots,\thermo{\Gamma}_K}^N$ are not invariant measures of the 2D Euler equations because finite-dimensional approximations of the Casimir invariants are not conserved quantities of the truncated dynamics \citep{Zeitlin1991}, the microcanonical measure $\mu_{\thermo{E_1},\thermo{E}_\infty,\thermo{L}_z,\thermo{\Gamma}_1,\ldots}$ is an invariant measure, thanks to the Liouville theorem (section~\ref{liouvillesection}) and to the fact that the quantities $\thermo{E_1},\thermo{E}_\infty,\thermo{L}_z,\thermo{\Gamma}_1,\ldots$ are dynamical invariants of the equation (section~\ref{eulerinvsection}).

Given the phase-space decomposition introduced above, the microcanonical measure could be expected to factor into a product of measures on $V_1$ and on $V_{\infty}$: $\mu_{\thermo{E_1},\thermo{E}_\infty,\thermo{L}_z,\thermo{\Gamma}_1,\ldots}(d\vort)=\mu_{\thermo{E}_1,\thermo{L}_z,\Gamma_{2,1}}(d\vort_1)\times \mu_{\thermo{E}_\infty,\Gamma_{k,\infty},\ldots}(d\vort_\infty)$. This is not true, because for $k\geq 3$, the Casimir invariants $\macro{G}_k[\vort_1+\vort_{\infty}]$ do not break up into separate functions of $\vort_1$ and $\vort_\infty$.

An interesting case occurs when we replace the full microcanonical measure $\mu_{\thermo{E_1},\thermo{E}_\infty,\thermo{L}_z,\thermo{\Gamma}_1,\ldots}$ by the \emph{energy-enstrophy measure}, which discards all the higher-order ($k \geq 3$) Casimir invariants:
\begin{align}
\mu_{\thermo{E_1},\thermo{E}_\infty,\thermo{L}_z,\thermo{\Gamma}_2}(d\vort)&=d\vort\frac{\delta(\macro{E}[\vort_1]-\thermo{E}_1)\delta(\macro{E}[\vort_{\infty}]-\thermo{E}_\infty) \delta(\macro{L}_z[\vort_1]-\thermo{L}_z) \delta(\macro{G}_2[\vort]-\thermo{\Gamma}_2)}{\Omega(\thermo{E_1},\thermo{E}_\infty,\thermo{L}_z,\thermo{\Gamma}_2)},
\end{align}
Note that there is no \emph{a priori} reason for discarding the higher-order Casimir invariants. Nevertheless, the general justification for the introduction of the energy-enstrophy measure is threefold. First, it is much more convenient to work with than the full microcanonical measure: in the context of the Robert-Sommeria-Miller (RSM) theory~\citep{Miller1990,Robert1991a,Robert1991b}, the energy-enstrophy microcanonical measure leads to a mean-field variational problem, the solutions of which form a sub-class of the solutions of the full RSM variational problem~\citep{Bouchet2008}. These solutions are characterized by a linear relationship between the vorticity and the stream function, which allows for analytical solutions of the mean field equation by decomposing the vorticity in terms of Laplacian eigenvectors~\citep{Chavanis1996a}. Besides, this sub-class of solutions already possess very interesting properties. From a practical point of view, they correspond to flow topologies which are indeed observed in real flows: monopoles, dipoles~\citep{Chavanis1996a} and Fofonoff flows~\citep{Venaille2011a,Naso2011} in a rectangular basin, solid-body rotations and dipoles on a sphere~\citep{Herbert2012a,Herbert2012b}, bottom-trapped currents in the ocean~\citep{Venaille2012b}... From a theoretical point of view, they exhibit peculiar thermodynamic properties, like bicritical points~\citep{Venaille2009}, second-order azeotropy~\citep{Venaille2011a}, marginal ensemble equivalence~\citep{Herbert2012a}... Finally, the energy-enstrophy variational problem has been connected to the full MRS variational problem in several limiting physical cases: the strong-mixing limit~\citep{Chavanis1996a} and the Gaussian small-scale vorticity prior~\citep{Ellis2000,Ellis2002,Chavanis2008b} for instance (see also the discussions in~\citep{Herbert2012a,Herbert2012b,Bouchet2012,CorvellecThesis}).

In addition to these general arguments, in the case of interest here, the energy-enstrophy measure has the convenient property of factoring out as a product of measures on $V_1$ and $V_\infty$, respectively:
\begin{align}
\mu_{\thermo{E_1},\thermo{E}_\infty,\thermo{L}_z,\thermo{\Gamma}_2}(d\vort)&=\mu_{\thermo{E_1},\thermo{L}_z,\thermo{\Gamma}_{21}}(d\vort_1)\mu_{\thermo{E}_\infty,\thermo{\Gamma}_{2\infty}}(d\vort_\infty),
\intertext{with}
\mu_{\thermo{E_1},\thermo{L}_z,\thermo{\Gamma}_{21}}(d\vort_1)&=d\vort_1\frac{\delta(\macro{E}[\vort_1]-\thermo{E}_1) \delta(\macro{L}_z[\vort_1]-\thermo{L}_z) \delta(\macro{G}_2[\vort_1]-\thermo{\Gamma}_{21})}{\Omega_1(\thermo{E_1},\thermo{L}_z,\thermo{\Gamma}_{21})},\\
\mu_{\thermo{E}_\infty,\thermo{\Gamma}_{2\infty}}(d\vort_\infty)&=d\vort_\infty\frac{\delta(\macro{E}[\vort_{\infty}]-\thermo{E}_\infty) \delta(\macro{G}_{2}[\vort_\infty]-\thermo{\Gamma}_{2\infty})}{\Omega_\infty(\thermo{E}_\infty,\thermo{\Gamma}_{2\infty})}.
\end{align}
Therefore, the random fields $\vort_1$ and $\vort_{\infty}$ can be considered as statistically independent in this case.
In the following section, we show that this allows for the construction of a simple mean-field theory, and we compute explicitly the statistical equilibria.

\textcolor{red}{Note that in fact, the dynamics on $V_1$ is integrable (see section \ref{integrabilitysection}). For a given value of the invariants $\thermo{E_1},\thermo{L}_z$, there is a single orbit $\vort_1^*(t)$ in $V_1$, described by \eqref{fundmodesdyneq}. This implies that the empirical measure $\mu_{\thermo{E_1},\thermo{L}_z}^*$, with density
\begin{align}
\rho_{\thermo{E_1},\thermo{L}_z}^*(\vort_1)&=\lim_{T \to +\infty} \int_0^T \delta(\vort_1^*(t)-\vort_1)dt,
\end{align}
and the energy-enstrophy microcanonical measure are proportional. In particular, the energy-enstrophy microcanonical measure on $V_1$ is ergodic.}

\subsection{Energy-enstrophy measure}\label{linearlimitsection}

\subsubsection{The energy-enstrophy mean-field theory}

The microcanonical measure can be transformed into a simpler expression, using large deviation properties, in the context of a mean-field theory referred to as the Robert-Sommeria-Miller theory. The underlying idea is that vorticity at different points can be treated as independent variables. The derivation of the mean-field theory, which goes beyond the scope if this paper, relies on large deviation properties for sets of measures known as Young measures~\citep{Robert1989,Michel1994b,Robert2000}. In this section, we just adapt to our case, in the energy-enstrophy limit, the mean-field theory as it was originally presented in the papers by Robert, Sommeria and Miller~\citep{Miller1990,Robert1991a,Robert1991b}. 

As $\vort_1(\vec{r})$ and $\vort_\infty (\vec{r})$ are independent random fields, we introduce the mean-field probability densities $\rho_1(\sigma_1,\vec{r})$ and $\rho_\infty(\sigma_\infty,\vec{r})$ such that
\begin{align}
\prob(\vort_1(\vec{r}) \in [\sigma_1-d\sigma,\sigma_1+d\sigma])&=\rho_1(\sigma_1,\vec{r})d\sigma,\\
\prob(\vort_\infty(\vec{r}) \in [\sigma_\infty-d\sigma,\sigma_\infty+d\sigma])&=\rho_\infty(\sigma_\infty,\vec{r})d\sigma.
\end{align}
The mean-field probability density for the vorticity to take on the value $\vort(\vec{r})=\vort_1(\vec{r})+\vort_\infty(\vec{r})$ at point $\vec{r}$ is given by the convolution
\begin{align}
\rho(\sigma,\vec{r})&=\int_\R d\sigma_1 \rho_1(\sigma_1,\vec{r}) \rho_\infty(\sigma- \sigma_1,\vec{r}).
\end{align}
The three probability densities $\rho_1,\rho_\infty,\rho$ are normalized in the following way:
\begin{align}
\forall \vec{r} \in \domain, \int_\R d\sigma_1 \rho_1(\sigma_1,\vec{r})=1, \quad \int_\R d\sigma_\infty \rho_\infty(\sigma_\infty,\vec{r})=1, \quad \int_\R d\sigma \rho(\sigma,\vec{r})=1.
\end{align}
We then define the coarse-grained vorticity fields
\begin{align}
\overline{\vort_1}(\vec{r})&=\int_{\R} \sigma_1 \rho_1(\sigma_1,\vec{r})d\sigma_1,\\
\overline{\vort_\infty}(\vec{r})&=\int_{\R} \sigma_\infty \rho_\infty(\sigma_\infty,\vec{r})d\sigma_\infty,\\
\overline{\vort}(\vec{r})&=\int_{\R} \sigma \rho(\sigma,\vec{r})d\sigma,
\intertext{and it is easily checked that $\overline{\vort}(\vec{r})=\overline{\vort}_1(\vec{r})+\overline{\vort}_\infty(\vec{r})$. We also define the mean-field energy, enstrophy, angular momentum and normalization functionals}
\meanfield{E}[\rho]&= \frac{1}{2}\int_\domain (\overline{\vort}-f)\overline{\psi}d\vec{r},\\
&=\int_\domain d\vec{r}\int_\domain d\vec{r'} \int_\R d\sigma \int_\R d\sigma' \sigma\sigma' H(\vec{r},\vec{r'})\rho(\sigma,\vec{r})\rho(\sigma',\vec{r'}),\\
&=\meanfield{E}[\rho_1]+\meanfield{E}[\rho_\infty],\\
\meanfield{G}_2[\rho] &= \int_{\domain} \overline{\vort^2}d\vec{r},\\
&= \int_{\domain} d\vec{r} \int_{\R} d\sigma \sigma^2 \rho(\sigma,\vec{r}),\\
&=\meanfield{G}_2[\rho_1]+\meanfield{G}_2[\rho_\infty],\\
\meanfield{L}_z[\rho] &=  \int_{\domain} d\vec{r} (\overline{\vort}-f) \cos \theta,\\
&= \int_{\domain} d\vec{r} \int_{\R}d\vort (\sigma-f) \cos \theta \rho(\sigma,\vec{r}),\\
&=\meanfield{L}_z[\rho_1],\\
\meanfield{N}[\rho](\vec{r}) &= \int_{\R} d\sigma \rho(\sigma,\vec{r}),\\
&=\meanfield{N}[\rho_1](\vec{r})\meanfield{N}[\rho_\infty](\vec{r}),
\intertext{where $H$ is (the opposite of) the Green function of the Laplacian:}
\psi(\vec{r})&= \int_\domain d\vec{r'} H(\vec{r},\vec{r'}) \omega(\vec{r'}).
\intertext{\textcolor{red}{The previous formulae are defined for any mean-field probability densities $\rho_1(\sigma_1,\vec{r}), \rho_\infty(\sigma_\infty,\vec{r})$. The usual argument of the Robert-Sommeria-Miller theory is a large deviation result which gives a variational principle determining the mean-field probability density, with a mean-field entropy which appears as a large-deviation rate. Here, the validity of the large deviation argument for $\rho_1(\sigma_1,\vec{r})$ is doubtful, because the underlying microcanonical measure is supported by a subspace whose dimension does not go to infinity, and the dynamics on this subspace is integrable. For now, let us suppose nevertheless that the mean-field probability density $\rho_1(\sigma_1,\vec{r})$ also satisfies a mean-field variational problem, and we will discuss the implications of the previous remark in section \ref{meanfieldfundamentalmodessection}.} The probability densities $\rho_1(\sigma_1,\vec{r}),\rho_\infty(\sigma_\infty,\vec{r})$ then maximize the mean-field entropies}
\meanfield{S}[\rho_1]&=-\int_{\domain} d\vec{r}\int_{\R} d\sigma_1 \rho(\sigma_1,\vec{r})\ln \rho(\sigma_1,\vec{r}),\\
\meanfield{S}[\rho_\infty]&=-\int_{\domain} d\vec{r}\int_{\R} d\sigma_\infty \rho(\sigma_\infty,\vec{r})\ln \rho(\sigma_\infty,\vec{r}),
\end{align}
so that the \emph{energy-enstrophy mean-field variational problem} reads
\begin{equation}
\begin{split}
\thermo{S}(\thermo{E}_1,\thermo{E}_\infty,\thermo{L}_z,\thermo{\Gamma}_{21},\thermo{\Gamma}_{2\infty})=\max_{\rho_1,\rho_\infty, \forall \vec{r} \in \domain, \meanfield{N}[\rho_1](\vec{r})=\meanfield{N}[\rho_\infty](\vec{r})=1} &\{ \meanfield{S}[\rho_1]+\meanfield{S}[\rho_\infty] | \meanfield{E}[\rho_1]=\thermo{E}_1, \meanfield{E}[\rho_\infty]=\thermo{E}_\infty,\\
&\meanfield{G}_{2}[\rho_1]=\thermo{\Gamma}_{21},\meanfield{G}_{2}[\rho_\infty]=\thermo{\Gamma}_{2\infty}, \meanfield{L}_z[\rho_1]=\thermo{L}_z \}.
\end{split}
\end{equation}

The critical points of this variational problem satisfy, for each independent perturbations $\delta\rho_1,\delta\rho_\infty$,
\begin{align}
&\begin{cases}
\delta \meanfield{S}[\rho_1]-\beta_1\delta \meanfield{E}[\rho_1]-\mu\delta\meanfield{L}_z[\rho_1]-\alpha_1\delta\meanfield{G}_2[\rho_1]=0\\
\delta \meanfield{S}[\rho_\infty]-\beta_\infty\delta \meanfield{E}[\rho_\infty]-\mu\delta\meanfield{L}_z[\rho_\infty]-\alpha_\infty\delta\meanfield{G}_2[\rho_\infty]=0
\end{cases}
\intertext{the solution of which is}
&\begin{cases}
\rho_1(\sigma_1,\vec{r})&=\frac{1}{\partfun_1} e^{-\beta_1 \sigma_1 \overline{\psi}_1(\vec{r})-\mu \sigma_1\cos \theta -\alpha_1\sigma_1^2}\\
\rho_\infty(\sigma_\infty,\vec{r})&=\frac{1}{\partfun_\infty} e^{-\beta_\infty \sigma_\infty \overline{\psi}_\infty(\vec{r}) -\alpha_\infty\sigma_\infty^2}
\end{cases}
\intertext{Straightforward computations yield}
\partfun_1&= \sqrt{\frac{\pi}{\alpha_1}}e^{\frac{(\beta_1\overline{\psi}_1(\vec{r})+\mu\cos\theta)^2}{4\alpha_1}}\\
\partfun_\infty&= \sqrt{\frac{\pi}{\alpha_\infty}}e^{\frac{(\beta_\infty\overline{\psi}_\infty(\vec{r}))^2}{4\alpha_\infty}}\\
\intertext{so that finally, we obtain the \emph{mean-field equation} characterizing statistical equilibria:}
\overline{\vort}_1&=-\frac{1}{\beta_1}\frac{\partial \ln \partfun_1}{\partial\psi_1}=\frac{\beta_1^2}{2\alpha_1}\overline{\psi}_1+\frac{\beta_1}{2\alpha_1}\mu\cos\theta\\
\overline{\vort}_\infty&=-\frac{1}{\beta_\infty}\frac{\partial \ln \partfun_\infty}{\partial\psi_\infty}=\frac{\beta_\infty^2}{2\alpha_\infty}\overline{\psi}_\infty.
\end{align}

\subsubsection{The coarse-grained variational problem and its critical points}

In this section, we show, following~\citep{Bouchet2008,Naso2010a}, that the mean-field equation characterizing the statistical equilibrium states of the 2D Euler equations on a rotating sphere in the limit of the energy-enstrophy measure can be recovered directly by the following variational problem, expressed directly in terms of the coarse-grained vorticity field:

\begin{subequations}
\begin{align}
\thermo{S}(\thermo{E}_1,\thermo{E}_\infty, \thermo{L}_z) &= \max_{\vort_1,\vort_{\infty}} \{ \macro{S}[\vort_1]+\macro{S}[\vort_{\infty}] | \macro{E}[\vort_1]=\thermo{E}_1, \macro{E}[\vort{_\infty}]=\thermo{E}_\infty, \macro{L}_z[\vort_1]=\thermo{L}_z\},\\
&= \max_{\vort_1} \{ \macro{S}[\vort_1] | \macro{E}[\vort_1]=\thermo{E}_1, \macro{L}_z[\vort_1]=\thermo{L}_z\} +  \max_{\vort_{\infty}} \{ \macro{S}[\vort_{\infty}] |  \macro{E}[\vort{_\infty}]=\thermo{E}_\infty\},
\end{align}
\end{subequations}
where $\macro{S}[\vort]=-\macro{G}_2[\vort]/2$ is the generalized entropy functional. This is the microcanonical variational problem. The relations of this variational problem with the mean-field variational problem have been discussed at length in previous publications~\citep{Bouchet2012,Herbert2012b}; the reader is referred to these articles for more information. The corresponding grand-canonical variational problem is
\begin{subequations}
\begin{align}
\thermo{J}(\beta_1,\beta_\infty, \mu) &= \max_{\vort_1,\vort_{\infty}} \{ \macro{S}[\vort_1]+\macro{S}[\vort_\infty] -\beta_1 \macro{E}[\vort_1] - \beta_\infty \macro{E}[\vort_{\infty}] - \mu \macro{L}_z[\vort_1]\},\\
&=\max_{\vort_1} \{ \macro{S}[\vort_1] -\beta_1 \macro{E}[\vort_1] - \mu \macro{L}_z[\vort_1]\} + \max_{\vort_{\infty}} \{ \macro{S}[\vort_\infty] - \beta_\infty \macro{E}[\vort_{\infty}] \}.
\end{align}
\end{subequations}

The critical points of the microcanonical and grand-canonical variational problems are the same, and they are given by
\begin{align}
&\delta\macro{S}[\vort_1]+\delta\macro{S}[\vort_\infty] -\beta_1 \delta\macro{E}[\vort_1] - \beta_\infty \delta\macro{E}[\vort_{\infty}] - \mu \delta\macro{L}_z[\vort_1]=0,\\
\intertext{and since the variations $\delta \vort_1, \delta\vort_{\infty}$ are independent, this amounts to the set of equations}
&\begin{cases}
\delta \macro{S}[\vort_1] -\beta_1 \delta\macro{E}[\vort_1] - \mu \delta\macro{L}_z[\vort_1]&=0,\\
\delta \macro{S}[\vort_\infty] - \beta_\infty \delta \macro{E}[\vort_{\infty}] &=0.
\end{cases}
\intertext{Clearly, this set of equation could be obtained directly from the separation into two independent contributions in the variational problem. Easy computations yield}
&\begin{cases}
\vort_1&= -\beta_1 \psi_1 - \mu \cos \theta,\\
\vort_\infty&= - \beta_\infty \psi_\infty,\\
\end{cases}
\intertext{or, equivalently, the mean-field equations}
&\begin{cases}
\Delta \psi_1 -\beta_1 \psi_1 &= f+ \mu \cos \theta,\\
\Delta \psi_\infty - \beta_\infty \psi_\infty&=0.\\
\end{cases}
\end{align}
The first equation was solved in~\citep{Herbert2012a,Herbert2012b} by discriminating cases where $\beta_1$ is an eigenvalue of the Laplacian and cases where it is not (see also~\cite{Chavanis1996a,Venaille2009,Venaille2011a,Naso2011} for applications of this general method). The general solution is of the form $\psi_1=3\thermo{L}_z/2\cos \theta+\sqrt{3[E_1-E_*(\thermo{L}_z)]}\sin \theta \cos (\phi-\phi_0)$ where the first term represents a solid-body rotation and the second a dipole flow. The amplitude of the dipole depends on the difference between the energy $E_1$ and the energy of a pure solid-body rotation with vertical angular momentum $\thermo{L}_z$, denoted $E_*(\thermo{L}_z)=3\thermo{L}_z^2/4$. The phase of the dipole $\phi_0$ is left undetermined by the conserved quantities. It was proved that this solution is stable in the microcanonical ensemble (and thus also in the grand-canonical ensemble); in other words, it is a maximum of the generalized entropy functional and not a minimum or a saddle-point.

The second equation can be solved using similar techniques --- it is even simpler because it is homogeneous. Clearly if $\beta_{\infty} \notin \Sp \Delta$, the mean-field equation has no solution. Now if $\beta_\infty=-\lambda_n$ (necessarily $n\geq 2$), the space of solutions is the corresponding eigenspace $\Ker (\Delta+\lambda_n \Id)$, that is to say, the shell $V_n$:
\begin{equation}
\psi_\infty = \sum_{m=-n}^n \psi_{nm} Y_{nm}.
\end{equation}
The solution space is thus degenerated $2n+1$ times.

\textcolor{red}{\subsubsection{Mean-field description of the fundamental modes and their integrable dynamics}\label{meanfieldfundamentalmodessection}}

\textcolor{red}{As mentioned above, the mean-field description for the fundamental modes $\vort_1(\vec{r})$ is questionable. In fact, it is not even necessary, because we know that the dynamics of this field is integrable. Instead of choosing the mean-field probability density maximizing the statistical entropy, one may consider the empirical density sampled by the integrable dynamics on the unique orbit defined by the invariants:
\begin{align}
\rho_1(\sigma_1,\vec{r})&=\lim_{T \to +\infty} \frac 1 T \int_0^T \delta(\sigma_1-\vort_1^*(\vec{r},t))dt,
\intertext{which gives a coarse-grained vorticity field $\overline{\vort}_1(\theta,\phi)=(3\thermo{L}_z+2\Omega)\cos \theta$. As expected, the dipole term does not survive time averaging. In fact, there is no need for this time averaging, and we may consider the time-varying mean-field probability density}
\rho_1(\sigma_1,\vec{r},t)&=\delta(\sigma_1-\vort_1^*(\vec{r},t)).
\end{align}
Of course, in this case the coarse-grained vorticity field is equal to the exact, fine-grained vorticity field: $\overline{\vort}_1(\theta,\phi,t)=\vort_1^*(\theta,\phi,t)$ defined by \eqref{fundmodesdyneq}. In particular, it corresponds to the statistical equilibrium found in the previous section, replacing the phase $\phi_0$ by $\phi_0+\Omega t$.}

\subsubsection{Stability of the critical points}

The stability for fundamental modes ($\psi_1$) as already been studied, both in the microcanonical and the grand-canonical ensembles~\citep{Herbert2012a}.
For the other modes ($\psi_\infty$), let us consider the free-energy functional $\macro{F}[\vort_\infty]=\macro{S}[\vort_\infty]-\beta_\infty \macro{E}[\vort_\infty]$. The stability condition in the microcanonical ensemble is $\delta^2 \macro{F}<0$ for every perturbation $\delta \vort_\infty \in V_\infty$ conserving the energy at first order. In the grand-canonical ensemble, this condition becomes $\delta^2 \macro{F}<0$ for every perturbation $\delta \vort_\infty \in V_\infty$.
Clearly, grand-canonical stability implies microcanonical stability, and microcanonical instability implies grand-canonical instability. Note that these thermodynamic stability criteria  also give necessary conditions for dynamic stability~\citep{Chavanis2009}. Here, we have,
\begin{equation}
\delta^2 \macro{F} = - \frac 1 2 \int_\domain \delta \vort_\infty^2 d\vec{r} - \frac {\beta_\infty} 2 \int_\domain (\nabla \delta \psi_\infty)^2 d\vec{r},
\end{equation}
or equivalently in spectral space, with the Plancherel theorem,
\begin{equation}\label{stabspeceq}
\delta^2 \macro{F} = - \frac 1 2 \sum_{n=2}^\infty \sum_{m=-n}^n \lambda_n (\lambda_n + \beta_\infty) | \delta \psi_{nm} |^2,
\end{equation}
with $\delta \psi_{nm} = \langle \delta \psi_\infty | Y_{nm} \rangle$. 
Clearly, if $\forall n \geq 2, \beta_\infty + \lambda_n >0$, then $\delta^2 \macro{F}<0$ (it is the Poincar\' e inequality). Thus, it suffices that $\beta_\infty > - \lambda_2$, to have $\delta^2 \macro{F} <0$. Here, this sufficient condition for thermodynamic stability is never satisfied, because the mean-field equation admits solution only for $\beta_\infty =-\lambda_n \leq -\lambda_2$ ($n \geq 2$).

Similarly to the global energy invariant case~\citep{Herbert2012b}, we can exhibit explicitly perturbations destabilizing the basic flow for $\beta_\infty<-\lambda_2$. Indeed, in this case it suffices to consider a perturbation $\delta\vort_\infty \in V_2$. This perturbation conserves the energy at first order: $\delta \macro{E}[\vort_\infty] = \langle \vort_\infty | \delta\psi_\infty \rangle/2+ \langle \delta \vort_\infty | \psi_\infty \rangle/2=0$ by orthogonality.
All the critical points with $\beta_\infty=-\lambda_n$ and $n\geq 3$ are thus saddle-points of the generalized entropy functional. These states are unstable in the microcanonical ensemble, and hence also in the grand-canonical ensemble.

It remains to treat the case $\beta_\infty = -\lambda_2$. In this case, the $n=2$ term vanishes in Eq. \ref{stabspeceq}. Mathematically, this means that the quadratic form $\delta^2 \macro{F}$ is \emph{degenerate}, and its radical is the subspace $\rad \delta^2 \macro{F} = V_2$. For every pertubation in $V_2$, $\delta^2 \macro{F}=0$. In this case, the free energy is zero: for $\vort_\infty \in V_2 $, $\mathcal{F}[\vort_\infty]=0$. 

In the microcanonical ensemble, all the states in the shell $V_2$ are metastable equilibria. There exist zero-energy perturbations (Goldstone modes) which allow to go from  one of these states to another one with the same energy. In the grand-canonical ensemble, the energy is not fixed: we can jump from any flow in the shell $V_2$ to any other flow in $V_2$. The situation is thus analogous to the case of the global energy invariant~\cite{Herbert2012a}: we have a manifold of metastable states ($V_2$ in the grand-canonical ensemble and the intersection between $V_2$ and the surface of energy $\macro{E}[\vort_\infty]$ in the microcanonical ensemble) with marginal ensemble equivalence.
The flows in the $V_2$ shell have the general form
\begin{align}
\psi_2&=\sum_{m=-2}^2 \psi_{2m} Y_{nm},
\intertext{where $\psi_{20}$ is real and $\psi_{2-1}=-\psi_{21}^*,\psi_{2-2}=\psi_{22}^*$ to ensure that the stream function takes on real values; or equivalently,}
\psi_2&= \tilde{\psi}_{20} (3\cos^2 \theta-1) + \tilde{\psi}_{21} \sin (2 \theta) \cos (\phi-\phi_1)+ \tilde{\psi}_{22} \sin^2 \theta \cos ( 2(\phi-\phi_2)),
\end{align}
where $\tilde{\psi}_{20}=\sqrt{\frac 5 {16\pi}}\psi_{20}, \tilde{\psi}_{21}=-\sqrt{\frac {15} {8\pi}}|\psi_{21}|, \tilde{\psi}_{22}=\sqrt{\frac {15} {8\pi}}|\psi_{22}|, \psi_{21}=|\psi_{21}|e^{-i\phi_1}, \psi_{22}=|\psi_{22}|e^{-2i\phi_2}$ and we have used the expression of the spherical harmonics of order 2 (see appendix~\ref{legpolformsection}).

\subsubsection{Thermodynamic Properties}

\begin{figure}
\centering
\includegraphics[width=0.7\textwidth]{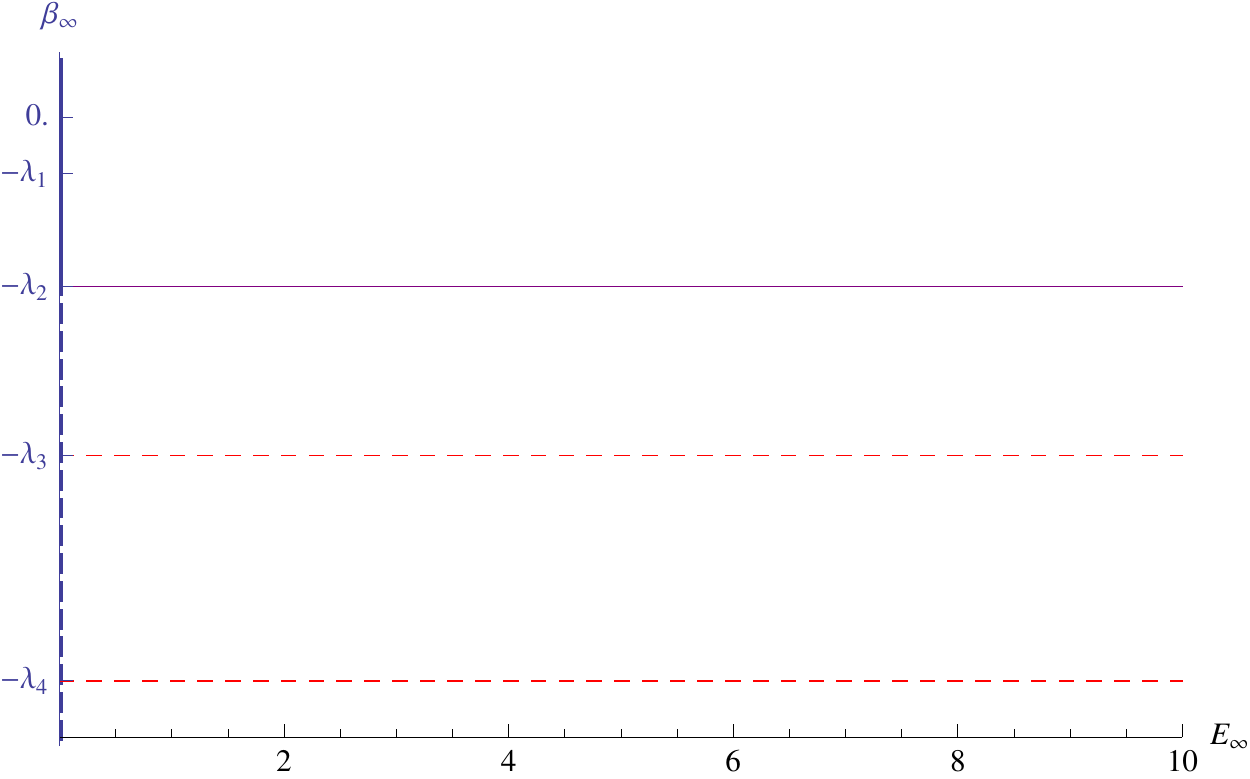}
\caption{Caloric curve $\thermo{E}_\infty(\beta_\infty)$. For $\thermo{E}_\infty>0$, the stable statistical equilibria (flows in the $V_2$ shell) are only reached for $\beta_\infty=-\lambda_2$ (purple solid line). For $\beta_\infty=-\lambda_n, n\geq 3$ (dashed horizontal red lines), we have saddle points of the entropy functional (flows in the $V_n$ shell), which are not thermodynamically stable. When $\thermo{E}_\infty=0$, the value of $\beta_\infty$ is not constrained (thick blue line); for $\beta_\infty \geq -\lambda_2$, the null flow is stable (solid blue line), otherwise it is only a saddle-point of the entropy functional (dashed blue line).}\label{caloriccurvefig}
\end{figure}

The caloric curve $E_1(\beta_1)$ and chemical potential curve $L(\mu)$ have already been published in~\citep{Herbert2012b}. The caloric curve $E_\infty(\beta_\infty)$ is also very simple: for $E_\infty=0$, $\beta_\infty$ can take on any value, while as soon as $E_\infty >0$, we must have $\beta_\infty=-\lambda_2$ (see Fig.~\ref{caloriccurvefig}).
The thermodynamic entropy is given by:
\begin{align}
\thermo{S}(\thermo{E}_1,\thermo{E}_\infty, \thermo{L}_z) &= \max_{\vort_1,\vort_{\infty}} \{ \macro{S}[\vort_1]+\macro{S}[\vort_{\infty}] | \macro{E}[\vort_1]=\thermo{E}_1, \macro{E}[\vort{_\infty}]=\thermo{E}_\infty, \macro{L}_z[\vort_1]=\thermo{L}_z\},\\
&= \max_{\vort_1} \{ \macro{S}[\vort_1] | \macro{E}[\vort_1]=\thermo{E}_1, \macro{L}_z[\vort_1]=\thermo{L}_z\} +  \max_{\vort_{\infty}} \{ \macro{S}[\vort_{\infty}] |  \macro{E}[\vort{_\infty}]=\thermo{E}_\infty\},\\
&=\thermo{S}_1(\thermo{E}_1,\thermo{L}_z)+\thermo{S}_\infty(\thermo{E}_\infty),\\
&=-\frac 2 3 \Omega^2 + \mu_c \thermo{L}_z -\lambda_1 \thermo{E}_1-\lambda_2\thermo{E}_\infty,
\end{align}
where $\thermo{S}_1(\thermo{E}_1,\thermo{L}_z)$ was already computed in~\cite{Herbert2012b}, with $\mu_c=-2\Omega$.
Analogously, the grand potential $\thermo{J}_\infty(\beta_\infty)$ vanishes, so that
\begin{align}
\thermo{J}(\beta_1,\beta_\infty, \mu) &= \max_{\vort_1,\vort_{\infty}} \{ \macro{S}[\vort_1]+\macro{S}[\vort_\infty] -\beta_1 \macro{E}[\vort_1] - \beta_\infty \macro{E}[\vort_{\infty}] - \mu \macro{L}_z[\vort_1]\},\\
&= \thermo{J}_1(\beta_1, \mu) + \thermo{J}_\infty(\beta_\infty),\\
&=\frac 1 3 \frac{(\mu-\mu_c)^2}{\beta_1-\lambda_1}-\frac 2 3 \Omega^2.
\end{align}

Note that the grand potential has a singularity at the phase transition point $(\beta_1,\mu)=(-\lambda_1,\mu_c)$. 
The thermodynamic parameters $\thermo{E}_1,\thermo{E}_\infty,\thermo{L}_z$ live in the region $\mathcal{T}$ of $\R^3$ defined by $\mathcal{T}=\{ (\thermo{E}_1,\thermo{E}_\infty,\thermo{L}_z) \in \R_+\times\R_+ \times \R, E_1\geq E_*(L) \}$. In the interior of $\mathcal{T}$, we recover the values of the Lagrange parameters through the usual thermodynamic identities:
\begin{align}
\beta_1&=\frac{\partial \thermo{S}(\thermo{E}_1,\thermo{E}_\infty, \thermo{L}_z)}{\partial \thermo{E}_1}, & \beta_\infty&=\frac{\partial \thermo{S}(\thermo{E}_1,\thermo{E}_\infty, \thermo{L}_z)}{\partial \thermo{E}_\infty}, & \mu&=\frac{\partial \thermo{S}(\thermo{E}_1,\thermo{E}_\infty, \thermo{L}_z)}{\partial \thermo{L}_z},\\
\beta_1&=-\lambda_1,		&	\beta_\infty&=-\lambda_2,			&	\mu&=\mu_c.
\end{align}
On the boundary of the thermodynamic domain $\partial\mathcal{T}$, the Lagrange parameters are not uniquely defined, only $\beta_1$ and $\mu$ being constrained by the relation $\frac{\mu-\mu_c}{\beta_1-\lambda_1}=constant$. In particular, the specific heats have a maximal discontinuity on $\partial\mathcal{T}$: they jump from $0$ (identically in $\mathring{\mathcal{T}}$) to $\infty$ on $\partial\mathcal{T}$.
This property is a consequence of the entropy surface $\thermo{S}(\thermo{E}_1,\thermo{E}_\infty, \thermo{L}_z)$ being a plane in thermodynamic space. In particular, it is a concave surface, but also a convex one: this is a situation of marginal ensemble equivalence at the thermodynamic level~\citep{Ellis2000,Touchette2004,Herbert2012a}.

\subsubsection{Conclusion}

The general form of the equilibrium states of the Euler equations on a rotating sphere in the energy-enstrophy limit is as follows
\begin{align}
\psi&=\psi_1+\psi_\infty,\\
\psi_1&=3\thermo{L}_z/2\cos \theta+\sqrt{3[E_1-E_*(\thermo{L}_z)]}\sin \theta \cos (\phi-\phi_0), \label{psionestateq}\\
\psi_\infty&=\tilde{\psi}_{20} (3\cos^2 \theta-1) + \tilde{\psi}_{21} \sin (2 \theta) \cos (\phi-\phi_1)+ \tilde{\psi}_{22} \sin^2 \theta \cos ( 2(\phi-\phi_2)),
\intertext{where the phases $\phi_0,\phi_1,\phi_2$ are left undetermined by the invariants of the Euler equations, and the amplitudes in $\psi_\infty$ are linked by the relation}
E_\infty&=\frac{16\pi}{5}\left\lbrack 3(\tilde{\psi}_{20})^2+(\tilde{\psi}_{21})^2+(\tilde{\psi}_{22})^2\right\rbrack.
\end{align}
Taking into account the integrable character of the dynamics in $V_1$,~\eqref{psionestateq} can be replaced by the time-dependent vorticity field~\eqref{fundmodesdyneq}.

\begin{figure}
\centering
\includegraphics[width=0.45\textwidth]{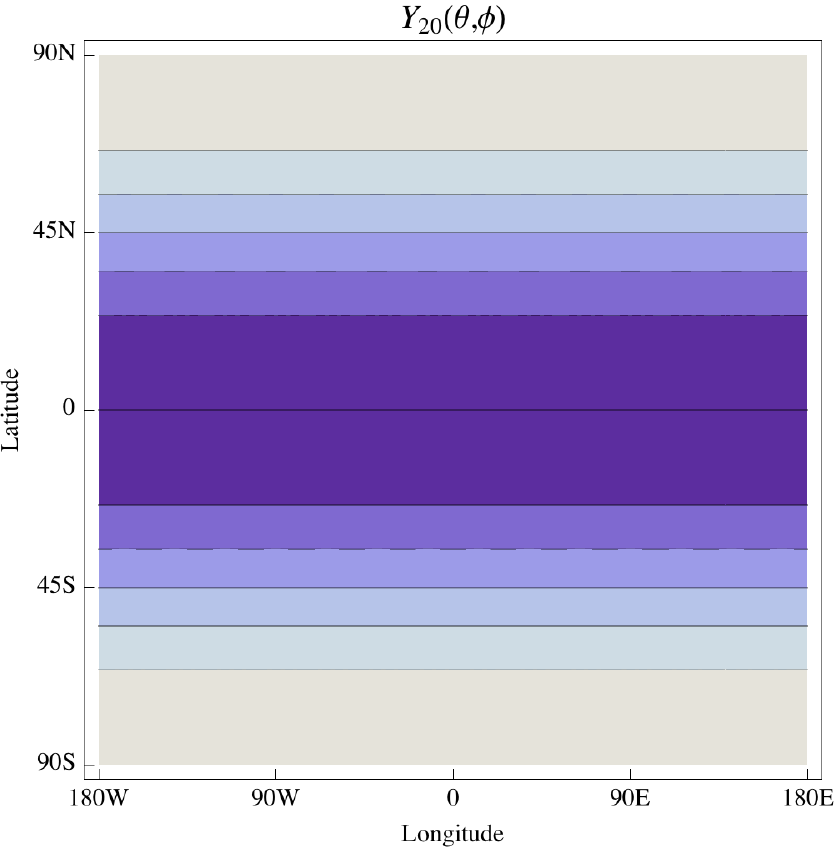}

\includegraphics[width=0.45\textwidth]{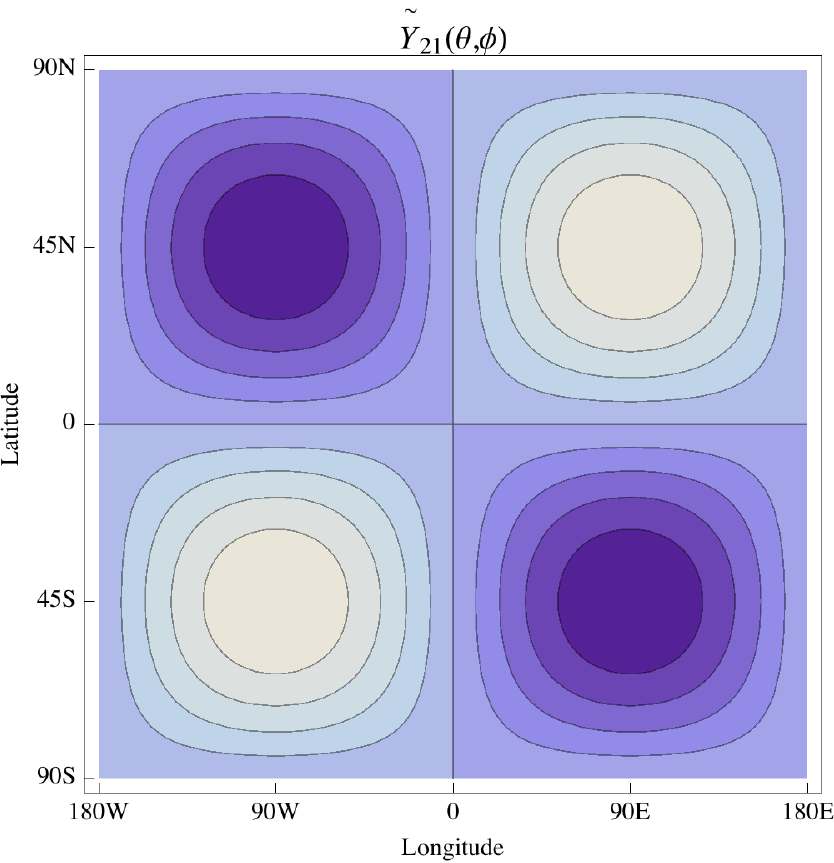}
\includegraphics[width=0.45\textwidth]{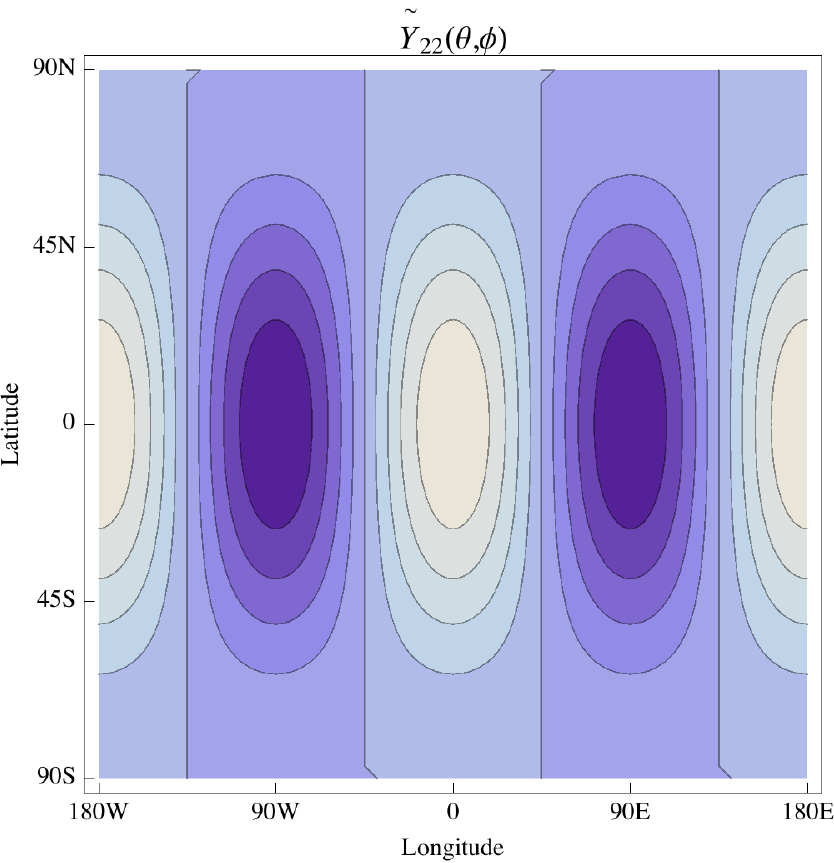}
\caption{Sample stream lines for flows in the shell $V_2$. There is one zonal mode with a velocity maximum at the equator, and two quadrupole modes with two vorticity maxima and two vorticity minima. The position of the vortices in the equilibrium flow depends on the specific mixture of these three modes, as well as on the background flow in $V_1$.}\label{quadrufig}
\end{figure}

In particular, $\psi_\infty \in V_2$, which means that all the energy $E_\infty$ condenses in the largest accessible scales (i.e. the gravest accessible modes).
The topology of the corresponding vorticity field $\vort_\infty$ is simple: when $\tilde{\psi}_{21}=\tilde{\psi}_{22}=0$, we have a zonal flow with \textcolor{red}{a velocity node at the equator and opposite sign velocities in each hemisphere}. Otherwise, the vorticity field has two maxima and two minima, the flow is thus dominated by four vortices, whose position depends on the value of the coefficients (see Fig.~\ref{quadrufig}).

\section{Discussion}

Our main result here is that due to the specificity of the spherical geometry considered here, the phase space breaks up into two invariant subspaces, which leads to independent conservation of two energy and two enstrophy invariants by the dynamics. \textcolor{red}{This property is also related to the precession of angular momentum in a rotating frame of reference, which means that the norm of the angular momentum is also a conserved quantity.} We have seen that taking into account these additional invariant, one may build a microcanonical measure, which can be simplified into an energy-enstrophy mean-field variational problem. The solutions of this variational problem can be analytically computed. We observe that the energy in excess of the fundamental modes condenses to the lowest accessible shell, $V_2$. When considering only the global energy invariant, condensation in $V_1$ was found~\citep{Herbert2012a}. Note that energy condensation is a very general feature of the energy-enstrophy measure, as shown by large deviation properties~\citep{Bouchet2010}. A consequence of the spherical geometry is therefore to halt the condensation process. This incomplete condensation occurs because, as we have shown, the interaction tensor blocks the energy transfers between the fundamental modes and the higher-order modes. Imposing rotation to the sphere is not sufficient to recover these transfers and allow full condensation of the energy; however, it is expected that interactions with a bottom topography would do so.
Note that the flow topology --- four vortices --- resulting from partial energy condensation coincides with stationary states observed in numerical simulations on a spherical domain~\citep{Cho1996,Marston2011}. It is sometimes argued that the presence of a quadrupole instead of a dipole is due to angular momentum conservation. It is true that on a non-rotating sphere, all the components of angular momentum are conserved. \textcolor{red}{In a rotating frame of reference, only the vertical component and the norm of the angular momentum are conserved. Nevertheless, we have shown that angular momentum has a precession motion in the rotating frame, which implies that the dynamics on the first shell $V_1$ is integrable and, in particular, independent of the other modes.} Hence, it is sufficient to say that due to the new invariants (angular momentum norm $L^2$ or, equivalently,  energy in the fundamental modes $E_1$), the excess energy $E-E_1$ will condense into the $V_2$ shell, which explains the formation of a quadrupole. Our statistical mechanics approach shows that even in the presence of rotation, quadrupole formation should occur in the limit of large times. Besides, due to the integrable dynamics in $V_1$, we expect the final state of the flow to be the superposition of a quadrupole and a time-dependent dipole, which behaves as a Rossby wave of unit wavenumber.

\appendix

\section{Some properties of the associated Legendre polynomials and Spherical Harmonics}\label{sphericalharmsection}

We recall in this appendix a few relations involving associated Legendre polynomials and spherical harmonics which are used in the paper. For more information, the reader is referred to \cite{GradshteynRyzhik,AbramowitzBook}.

\subsection{Definitions}

The \emph{Legendre polynomials} $(P_n)_{n \in \N}$ are a family of orthogonal polynomials: considering the scalar product (on $\R[X]$) $( P, Q )=\int_{-1}^1 P(x)Q(x)dx$, they can be obtained thanks to the Gram-Schmidt orthogonalization process starting from the canonical basis $(1,X,\ldots,X^n,\ldots)$. As usual for families of orthogonal polynomials, they also satisfy
\begin{enumerate}[label=(\roman*)]
\item The recursion formula $(n+1)P_{n+1}(x)-(2n+1)xP_n(x)-nP_{n-1}(x)=0$.
\item The differential equation $(1-x^2)P_n''(x)-2xP_n'(x)+n(n+1)P_n(x)=0$.
\item The Rodrigues formula $P_n(x)=\frac 1 {2^n n!} \frac{d^n}{dx^n} \left\lbrack(x^2-1)^n\right\rbrack$.
\end{enumerate}

The associated Legendre polynomials can be defined from the Legendre polynomials as follows, for $-n \leq m \leq n$:
\begin{equation}
P_n^m(x)=(-1)^m (1-x^2)^{\frac m 2} \frac {d^m P_n(x)}{dx^m}.
\end{equation}
The associated Legendre polynomials are also solutions of the following differential equation:
\begin{equation}
(1-x^2)y''-2xy'+\left\lbrack n(n+1) - \frac{m^2}{1-x^2}\right\rbrack y=0,
\end{equation}
and they satisfy a relation analogous to the Rodrigues formula:
\begin{equation}\label{rodrigueseq}
P_n^m(x)=\frac {(-1)^m} {2^n n!} (1-x^2)^{\frac m 2} \frac{d^{n+m}}{dx^{n+m}} \left\lbrack(x^2-1)^n\right\rbrack.
\end{equation}

We introduced earlier the spherical harmonics as the eigenfunctions of the Laplacian. They form an orthonormal basis of the Hilbert space $L^2(\sphere)$. They are given by the expression:
\begin{align}
Y_{nm}(\theta,\phi)&= \sqrt{\frac{2n+1}{4\pi}\frac{(n-m)!}{(n+m)!}} P_m^m(\cos \theta) e^{im\phi},
\end{align}
and are eigenfunctions of the Laplacian with eigenvalue $-\beta_n=-n(n+1), \Delta Y_{nm}=-\beta_n Y_{nm}$, where the Laplacian on the sphere $\sphere$ is given by
\begin{equation}
\Delta f= \frac{1}{\sin \theta} \frac{\partial}{\partial \theta} \left( \sin \theta\frac{\partial f}{\partial \theta}\right)+\frac{1}{\sin^2	 \theta} \frac{\partial^2 f}{\partial \phi^2}.
\end{equation}

\subsection{Some general properties}\label{legpolrelsection}

The associated Legendre polynomials and spherical harmonics satisfy the following relations ($n \in \N, -n \leq m \leq n$):
\begin{align}
P_n^{-m}(x) &= (-1)^m \frac{(n-m)!}{(n+m)!} P_n^m(x),\\
Y_{nm}(\theta,\phi)^*&=(-1)^m Y_{n,-m}(\theta,\phi).
\end{align}
With standard normalizations, the orthogonality properties of the associated Legendre polynomials and spherical harmonics are:
\begin{align}
\int_{-1}^1 P_n^m(x)P_p^m(x)dx&=\frac{2(n+m)!}{(2n+1)(n-m)!}\delta_{np},\\
\int_{-1}^1 \frac{P_n^m(x)P_n^q(x)}{1-x^2}dx&= \frac{(n+m)!}{m(n-m)!}\delta_{mq}, \qquad (m,q)\neq (0,0),\\
\int_{\sphere} Y_{nm}(\theta,\psi)Y_{pq}(\theta,\psi)^*d\Omega &=\delta_{np}\delta_{mq}.
\end{align}

A certain number of formulae (see \cite{GradshteynRyzhik}) relate products of associated Legendre polynomials (or their derivative) with simple quantities such as $x,\sqrt{1-x^2},(1-x^2)$ (which are actually low-order associated Legendre polynomials, see section \ref{legpolformsection}). The only one that we use in this paper is the following:
\begin{equation}
(x^2-1){P_n^m}'(x) = -(n+m)(n-m+1)\sqrt{1-x^2} P_n^{m-1}(x) -mxP_n^m(x).
\end{equation}

Note that although they are called \enquote{polynomials}, the associated Legendre polynomials $P_n^m$ are polynomials (with degree $n$) only when $m$ is even. When $m$ is odd, they take the form of a product of $\sqrt{1-x^2}$ and a polynomial (with degree $n-1$).
From the Rodrigues formula (Eq. \ref{rodrigueseq}), it is clear that as soon as $m \neq 0$, $1$ and $-1$ are both roots of $P_n^m$.

\subsection{The first few associated Legendre polynomials and spherical harmonics}\label{legpolformsection}

Here is the explicit expression for the associated Legendre polynomials and spherical harmonics up to $n=2$.

\begin{align*}
P_0^0(x)		&=1					&	Y_{00}(\theta,\phi)	&=\sqrt{\frac{1}{4\pi}}								\\
P_1^{-1}(x)	&=\frac 1 2 \sqrt{1-x^2}	&	Y_{1-1}(\theta,\phi)	&=\sqrt{\frac{3}{8\pi}}\sin \theta e^{-i\phi}				\\
P_1^0(x)		&=x					&	Y_{10}(\theta,\phi)	&=\sqrt{\frac{3}{4\pi}}\cos \theta						\\
P_1^1(x)		&=-\sqrt{1-x^2}			&	Y_{11}(\theta,\phi)	&=-\sqrt{\frac{3}{8\pi}}\sin \theta e^{i\phi}				\\
P_2^{-2}(x)	&=\frac 1 8 (1-x^2)		&	Y_{2-2}(\theta,\phi)	&=\sqrt{\frac{15}{32\pi}}\sin^2\theta e^{-2i\phi}			\\
P_2^{-1}(x)	&=\frac x 2 \sqrt{1-x^2}	&	Y_{2-1}(\theta,\phi)	&=\sqrt{\frac{15}{8\pi}}\sin\theta\cos\theta e^{-i\phi}		\\
P_2^0(x)		&=\frac 1 2 (3x^2-1)		&	Y_{20}(\theta,\phi)	&=\sqrt{\frac{5}{16\pi}}(3\cos^2\theta-1)				\\
P_2^1(x)		&=-3x\sqrt{1-x^2}		&	Y_{21}(\theta,\phi)	&=-\sqrt{\frac{15}{8\pi}}\sin\theta\cos\theta e^{i\phi}		\\
P_2^2(x)		&=3(1-x^2)			&	Y_{22}(\theta,\phi)	&=\sqrt{\frac{15}{32\pi}}\sin^2\theta e^{2i\phi}			\\
\end{align*}

\section{Direct computations on the interaction tensor}\label{directcompusection}

\textcolor{green}{\subsection{Fundamental-mode excitation}\label{fundmodeexcsec}
In this section, we show that fundamental modes --- $n=1$ --- cannot be excited by quadratic interactions. In other words, $B^{1m}_{n_1m_1n_2m_2} = 0$. Taking into account the results of the previous section, it suffices to prove this for $n_1 \neq n_2$ and $(m_1,m_2)\neq (0,0)$. Due to the presence of a Dirac delta factor $\delta_{m_1+m_2,m}$ in~\eqref{Atensoreq}, we can also suppose $m_1+m_2=m$ without loss of generality, and prove that $m_1 I_{nmn_1m_1}^{n_2m_2}=m_2 I_{nmn_2m_2}^{n_1m_1}$. As $-1 \leq m \leq 1$, there are only three cases.
\paragraph{Case $m=0$.}
Integrating by parts,
\begin{subequations}
\begin{align}
I_{10n_1m_1}^{n_2m_2} &= \int_{-1}^1 x P_{n_1}^{m_1}(x) {P_{n_2}^{m_2}}' (x) dx,\\
&= \left\lbrack x P_{n_1}^{m_1}(x) {P_{n_2}^{m_2}}(x)  \right\rbrack_{-1}^1 - \int_{-1}^1 P_{n_1}^{m_1}(x) {P_{n_2}^{m_2}}(x)dx - \int_{-1}^1 x {P_{n_1}^{m_1}}'(x) {P_{n_2}^{m_2}}(x) dx.
\end{align}
\end{subequations}
As $m_1$ cannot be zero, $1$ and $-1$ are both roots of $P_{n_1}^{m_1}$ and the brackets vanish. Likewise, the second term vanishes thanks to an orthogonality property, because $P_{n_2}^{m_2}(x)=P_{n_2}^{-m_1}(x)= (-1)^{m_1} \frac {(n_2-m_1)!} {(n_2+m_1)!}P_{n_2}^{m_1}(x)$ and since $n_1 \neq n_2$, $P_{n_1}^{m_1}$ and $P_{n_2}^{m_1}$ are orthogonal (see appendix~\ref{legpolrelsection}). Replacing the third term by the definition of the structure integrals yields
\begin{equation}
I_{10n_1m_1}^{n_2m_2}+ I_{10n_2m_2}^{n_1m_1}=0,
\end{equation}
which is exactly the result we wanted to prove.
\paragraph{Case $m=1$.}
Now we must have $m_1+m_2=1$, so that we need to prove that
\begin{equation}
m_1 \left( I_{11n_1m_1}^{n_2m_2} + I_{11n_2m_2}^{n_1m_1} \right) = I_{11n_2m_2}^{n_1m_1}.
\end{equation}
Again, integrating by parts yields
\begin{subequations}
\begin{align}
I_{11n_1m_1}^{n_2m_2} &= - \int_{-1}^1 \sqrt{1-x^2} P_{n_1}^{m_1}(x) {P_{n_2}^{m_2}}'(x) dx,\\
&=- \left \lbrack \sqrt{1-x^2} P_{n_1}^{m_1}(x) P_{n_2}^{m_2}(x) \right \rbrack_{-1}^1 + \int_{-1}^1 \sqrt{1-x^2} {P_{n_1}^{m_1}}'(x) P_{n_2}^{m_2}(x) - \int_{-1}^1 x P_{n_1}^{m_1}(x) P_{n_2}^{m_2}(x) \frac {dx} {\sqrt{1-x^2}}.
\end{align}
\end{subequations}
Since $m_1$ and $m_2$ cannot vanish simultaneously, $1$ and $-1$ are both roots of (at least) one of the Legendre associated functions in the brackets, which vanishes. The second term is equal to $-I_{11n_2m_2}^{n_1m_1}$, so that
\begin{equation}
m_1 \left( I_{11n_1m_1}^{n_2m_2} + I_{11n_2m_2}^{n_1m_1} \right) = - \int_{-1}^1 m_1 x P_{n_1}^{m_1}(x) P_{n_2}^{m_2}(x) \frac {dx} {\sqrt{1-x^2}}.
\end{equation}
Using the following formula (see appendix~\ref{legpolrelsection}) for Legendre associated functions \citep{GradshteynRyzhik}:
\begin{equation}
(x^2-1){P_n^m}'(x) = -(n+m)(n-m+1)\sqrt{1-x^2} P_n^{m-1}(x) -mxP_n^m(x),
\end{equation}
in the right-hand side, we obtain
\begin{equation}
m_1 \left( I_{11n_1m_1}^{n_2m_2} + I_{11n_2m_2}^{n_1m_1} \right) = I_{11n_2m_2}^{n_1m_1} + (n_1+m_1)(n_1-m_1+1) \int_{-1}^1 P_{n_1}^{m_1-1}(x) P_{n_2}^{m_2}(x) dx.
\end{equation}
The second term of the right-hand side vanishes because $P_{n_1}^{m_1-1}=P_{n_1}^{-m_2}$, which is proportional to $P_{n_1}^{m_2}$ and since $n_1 \neq n_2$ the integral vanishes by orthogonality of the Legendre polynomials. The expected result follows.
\paragraph{Case $m=-1$.}
Using the proportionality relation between $P_n^m$ and $P_n^{-m}$, we obtain the relations
\begin{subequations}
\begin{align}
m_1 I_{n,-m,n_1m_1}^{n_2m_2}&=(-1)^m \frac{(n-m)!}{(n+m)!} m_1 I_{nmn_1m_1}^{n_2m_2},\\
m_2 I_{n,-m,n_2m_2}^{n_1m_1}&=(-1)^m \frac{(n-m)!}{(n+m)!} m_2 I_{nmn_2m_2}^{n_1m_1},
\end{align}
\end{subequations}
so that for $n=m=1$, this case immediately follows from the previous one.}

\textcolor{green}{\subsection{Fundamental-mode interactions}\label{fundmodeintersec}
It is already clear from the general considerations above (see section~\ref{interactgensection}) that the quadratic interaction of two fundamental modes --- $n_1=n_2=1$ --- have a vanishing contribution to any other mode: $B^{nm}_{1m_11m_2} =0$. However, we do not know if $A^{nm}_{1m_11m_2} =0$, which has an impact on how the fundamental modes of the flow interact with rotation, through the linear term $\mathcal{L}_{n,m}[\vort]$. In fact, if $n>1, A^{nm}_{1m_11m_2} =0$, as we shall prove in this section.
\paragraph{Case $m_1=0$.} From~\eqref{Atensoreq}, it suffices to prove that $I^{10}_{nm1m_2}=0$. As $P_{1}^{0}(x)=x$, we have
\begin{align}
I^{10}_{nm1m_2}&=\int_{-1}^1 P_n^m(x) P_{1}^{m_2}(x) dx,
\end{align}
but due to the delta factor, we can assume $m=m_2$, and since $n>1$, this integral vanishes by orthogonality of the Legendre associated polynomials.
By symmetry, this also solves the case $m_2=0$.
\paragraph{Case $m_1=1$.} The case $m_2=0$ was treated above and we already know that the result holds when $m_1=m_2$, hence we can assume without loss of generality that $m_2=-1$ and $m=0$. Note that
\begin{align}
P_1^1(x)&=-\sqrt{1-x^2}, & {P_1^1}'(x)&=\frac{x}{\sqrt{1-x^2}},\\
P_1^{-1}(x)&=\frac 1 2\sqrt{1-x^2}, & {P_1^{-1}}'(x)&=-\frac{x}{2\sqrt{1-x^2}},
\end{align}
so that
\begin{align}
I_{nm1-1}^{11}=I_{nm11}^{1-1}&=\frac 1 2 \int_{-1}^1 x P_n^m(x)dx\\
&=\frac 1 2 \int_{-1}^1 P_1^0(x) P_n^0(x)dx\\
&=\frac 1 2 \delta_{1n}
\end{align}
Therefore, for $n>1, I_{nm1-1}^{11}=I_{nm11}^{1-1}=0$ and $A^{nm}_{111-1} =0$. 
Due to the symmetry properties, we have proved that $A^{nm}_{1m_11m_2} =0$ for $-1\leq m_1,m_2\leq 1$ and $n>1$.}

\subsection{Direct proof for the conservation of $\macro{E}[\vort_1],\macro{E}[\vort_\infty]$}\label{energyinvdirproofsection}

The energy in shell $V_n$ is given by
\begin{align}
\macro{E}[\vort_n]&=\frac 1 2 \sum_{m=-n}^n (\vort_{nm}-\delta_{n1}\delta_{m0}f_{10})\psi_{nm}^* = \frac {\lambda_n} 2 \sum_{m=-n}^n | \psi_{nm} |^2,
\intertext{so that the derivative is}
\dot{\macro{E}}[\vort_n]&=-\frac {1} 2 \sum_{m=-n}^n  \psi_{nm}^* \mathcal{Q}_{nm}[\vort] + \frac 1 2 \sum_{m=-n}^n \psi_{nm}^* \mathcal{L}_{nm}[\vort] + cc,
\intertext{where the star denotes complex conjugacy and $cc$ is the complex conjugate of the whole expression. In particular, for $n=1$, we have proved (section~\ref{fundmodeexcsec}) that $\mathcal{Q}_{1m}[\vort]  = 0$ and $\mathcal{L}_{1m}[\vort]=\mathcal{L}_{1m}[\vort_1]$. Hence,}
\dot{\macro{E}}[\vort_1]&=\frac 1 2 \sum_{m=-1}^1 \psi_{1m}^* \sum_{m'=-1}^1 \vort_{1m'} \frac{A_{101m'}^{1m}}{\lambda_1}f_{10} + cc,\\
&=\frac 1 2 \sum_{m=-1}^1 \psi_{1m}^* \vort_{1m} \frac{A_{101m}^{1m}}{\lambda_1}f_{10} + cc,
\intertext{because of the $\delta_{m,m'}$ factor contained in $A_{101m}^{1m'}$, and since $A_{101m}^{1m}$ vanishes for $m=0$,}
&=\frac 1 2 \sum_{m=-1}^1 \psi_{1m}^* \psi_{1m} A_{101m}^{1m}f_{10} + cc,\\
&=\frac 1 2 \sum_{m=-1}^1 f_{10} |\psi_{1m}|^2 (A_{101m}^{1m} + cc),\\
&=0,
\end{align}
because $A_{101m}^{1m}$ is an imaginary number.
As $\macro{E}[\vort]$ is conserved by the standard arguments, it follows that $\macro{E}[\vort_{\infty}]$ is also conserved. A similar result holds for the enstrophy.

\begin{acknowledgements}
The author acknowledges the \emph{Laboratoire des Sciences du Climat et de l'Environnement} and \emph{Service de Physique de l'Etat Condens\'e}, CEA Saclay, Gif-sur-Yvette, France, where part of this work was done. The National Center for Atmospheric Research is sponsored by the National Science Foundation.
\end{acknowledgements}

\bibliographystyle{spmpsci}      

\end{document}